\documentclass[iop]{emulateapj}
\usepackage{hyperref}
\usepackage{longtable}

\usepackage{natbib}
\shorttitle{Magnetic Fields In Relativistic Collisionless Shocks}
\shortauthors{Santana, Barniol Duran, \& Kumar}

\begin{document}

\title{Magnetic Fields In Relativistic Collisionless Shocks}

\author{Rodolfo Santana\altaffilmark{1a}, Rodolfo Barniol Duran\altaffilmark{2b}, \& Pawan Kumar\altaffilmark{1c}}
\altaffiltext{1}{Department of Astronomy, University of Texas at Austin, Austin, TX 78712, USA}
\altaffiltext{2}{Racah Institute for Physics, The Hebrew University, Jerusalem, 91904, Israel}
\email{(a) santana@astro.as.utexas.edu; (b) rbarniol@phys.huji.ac.il} 
\email{(c) pk@astro.as.utexas.edu}

\begin{abstract}
We present a systematic study on 
magnetic fields in Gamma-Ray Burst (GRB) external forward shocks (FSs). 
There are 60 (35) GRBs in our X-ray (optical) sample, mostly from
\textit{Swift}. We use two methods to study $\epsilon_{B}$ (fraction of energy
in magnetic field in the FS). 1. For the X-ray sample, 
we use the constraint that the observed flux at the end of the
steep decline is $\ge$ X-ray FS flux. 2. For the optical sample, we use the condition that the 
observed flux arises from the FS (optical sample light curves 
decline as $\sim t^{-1}$, as expected for the FS). 
Making a reasonable assumption on $E$ (jet isotropic equivalent 
kinetic energy), we converted these conditions into an upper limit 
(measurement) on $\epsilon_{B} n^{2/(p+1)}$ for our X-ray (optical)
sample, where $n$ is the circumburst density and $p$ is the electron 
index. Taking $n=1 \mbox{ cm}^{-3}$, the 
distribution of $\epsilon_{B}$ measurements (upper limits) for our 
optical (X-ray) sample has a range of $\sim 10^{-8} -10^{-3}$ 
($\sim 10^{-6} -10^{-3}$) and median of 
$\sim \mbox{ few} \times 10^{-5}$ ($\sim \mbox{ few} \times 10^{-5}$).
To characterize how much amplification is needed, beyond shock compression of a seed
magnetic field $\sim 10 \mu \mbox{G}$, we expressed our results in
terms of an amplification factor, $AF$, which is very weakly dependent 
on $n$ ($AF \propto n^{0.21}$). The range of $AF$
measurements (upper limits) for our optical (X-ray) sample is 
$\sim 1-1000$ ($\sim 10-300$) with a median of $\sim 50$ ($\sim 50$). 
These results suggest that some amplification, in addition to shock
compression, is needed to explain the afterglow
observations.
\end{abstract}


\maketitle

\section{Introduction}

Gamma-Ray Bursts (GRBs) are bright explosions occurring at cosmological
distances which release
gamma-rays for a brief time, typically on a timescale of $\sim \mbox{
  few} \times$ 10 sec
(e.g. \citealt{piran_2004}, \citealt{gehrels_et_al_2009}, \citealt{zhang_2011}). This short-lived emission 
of gamma-rays is known as the prompt emission. After the prompt 
emission, long-lived emission in the X-ray, optical, and radio bands (on
timescales of days, months, or even years) is also observed from what
is called the ``afterglow''. Although the mechanism for the
prompt emission is currently being debated, the afterglow
emission has a well-established model based on external shocks 
\citep[][]{rees_and_meszaros_1992,meszaros_and_rees_1993,paczynski_and_rhoads_1993}. In this 
framework, a relativistic jet emitted by the central engine interacts with the medium surrounding 
the GRB progenitor. This interaction produces two shocks; the external-reverse shock
and the external-forward shock \citep{meszaros_and_rees_1997,sari_and_piran_1999}.
The external-reverse shock heats up the jet while the external-forward
shock heats up the
medium surrounding the explosion. The external-reverse shock 
is believed to be short-lived in the optical band
\citep{sari_and_piran_1999a} and might have been observed, perhaps, in
a few cases. The long-lived afterglow emission is
interpreted as synchrotron radiation from the external-forward shock. This
shock is taken to produce a power-law distribution of high energy electrons 
and to amplify the pre-existing seed magnetic field in the surrounding medium. 
These high energy electrons are then accelerated by the amplified magnetic field and emit radiation
by the synchrotron process.

One of the open questions in the field of GRB afterglows is: what is the dynamo
mechanism amplifying magnetic fields in the collisionless relativistic shocks involved for GRB external shocks? 
The magnetic field strength downstream of the shock front is expressed in
terms of $\epsilon_{B}$, which is defined as the fraction of energy
that is in the magnetic field downstream of the shock front. With
this definition, the explicit expression for $\epsilon_{B}$ is
\begin{equation}
\epsilon_{B} = \frac{B^{2}}{32 \pi n m_{p} c^{2} \Gamma^{2}} ,
\end{equation}
where $B$ is the co-moving magnetic field downstream of the shock front, $n$ is the
density surrounding the GRB progenitor, $m_{p}$ is the proton mass,
$c$ is the speed of light, and $\Gamma$ is the Lorentz factor of the
shocked fluid downstream of the shock front 
\citep[e.g.][]{sari_et_al_1998,wijers_and_galama_1999,panaitescu_and_kumar_2000}. If shock compression is the
only mechanism amplifying the magnetic field downstream of the shock
front, then $B$ is given by $B = 4 \Gamma B_{0}$ \citep[e.g.][]{achterberg_et_al_2001}, where $B_{0}$ is the
seed magnetic field in the medium surrounding the GRB
progenitor. Using this expression for $B$, $\epsilon_{B}$ simplifies
to $\epsilon_{B} = B_{0} ^{2} / 2 \pi n m_{p} c^{2}$. Using the
value for the ambient magnetic field of the Milky Way galaxy
$B_{0} \sim \mbox{ few } \mu \mbox{G}$ and a density for the
surrounding medium of $n = 1 \mbox{ cm}^{-3}$, $\epsilon_{B}$ is
expected to be $\sim 10^{-9}$.

Several studies have modeled afterglow data to determine what values of the afterglow parameters best describe
the observations
\citep[e.g.][]{wijers_and_galama_1999,panaitescu_and_kumar_2002,yost_et_al_2003,panaitescu_2005}. 
The results from previous studies show that $\epsilon_{B}$ 
ranges from $\epsilon_{B} \sim 10^{-5}-10^{-1}$. These values for
$\epsilon_{B}$ are much larger than the $\epsilon_{B} \sim
10^{-9}$ expected from shock compression alone and suggest that some
additional amplification is needed to explain the observations. There have been
several theoretical and numerical studies that have considered possible mechanisms,
operating in the plasma in the medium surrounding the GRB, that can generate extra amplification for
the magnetic field. The mechanisms that have been proposed are the
two-stream Weibel instability 
\citep[][]{weibel_1959,medvedev_and_loeb_1999,gruzinov_and_waxman_1999,silva_et_al_2003,medvedev_et_al_2005} 
and dynamo generated by turbulence 
\citep[][]{milosavljevic_and_nakar_2006,milosavljevic_et_al_2007,sironi_and_goodman_2007,goodman_and_macfadyen_2008,couch_et_al_2008,zhang_macfadyen_et_al_2009,mizuno_et_al_2011,inoue_et_al_2011}.

Recent results \citep[][]{kumar_and_barniol_duran_2009,kumar_and_barniol_duran_2010} 
found surprisingly small values of $\epsilon_{B} \sim 10^{-7}$
for 3 bright GRBs with \textit{Fermi}$/$LAT detections. These
values of $\epsilon_{B}$ are $\sim 2$ orders of magnitude smaller than
the smallest previously reported $\epsilon_{B}$ value and they can be explained with the only
amplification coming from shock compression of a seed magnetic
field of a few $10 \mu \mbox{G}$ \footnote{The values given above of
  $\epsilon_{B} \sim 10^{-7}$ are under the assumption of 
$n  = 1 \mbox{ cm}^{-3}$. It is important to note that when reaching
the conclusion that shock compression provides enough amplification to
explain the afterglow data, 
\citet[][]{kumar_and_barniol_duran_2009,kumar_and_barniol_duran_2010}
did not assume a value for $n$. Also, the results of small
$\epsilon_{B} \sim 10^{-7}$ values do not depend on whether or not the LAT emission is produced by the external-forward
shock. These small $\epsilon_{B}$ values were inferred from the late time X-ray and optical data and from the
constraint that the external-forward shock does not produce flux at
150 keV that exceeds the observed prompt emission flux at 50 seconds.}. Although this seed magnetic field
is stronger than the one of the Milky Way galaxy by about
a factor $\sim 10$, seed magnetic fields of a few $10 \mu \mbox{G}$ have
been measured before. The seed magnetic fields in the spiral arms of some
gas-rich spiral galaxies with high star formation rates have been measured
to be 20-30 $\mu \mbox{G}$ \citep{beck_2011}. Seed magnetic fields as high
as $0.5-18$ mG were measured in starburst galaxies by measuring the Zeeman
splitting of the OH megamaser emission line at 1667 MHz \citep{robishaw_et_al_2008}.

Given this disagreement between the recent and previous results,
the question regarding the amplification of magnetic fields in GRB external
relativistic collisionless shocks remains unanswered. The first goal of this
study is to provide a systematic determination of $\epsilon_{B}$ for a
large sample of GRBs by using the same method to determine
$\epsilon_{B}$ for each burst in our X-ray or optical sample. This is the first time
such a large and systematic study has been carried out for 
$\epsilon_{B}$. Knowing the value of $\epsilon_{B}$ for large samples
will help us determine how much amplification of the magnetic field 
is needed to explain the afterglow observations. We mostly limit our samples to GRBs detected by the 
\textit{Swift} satellite, with measured redshift. 
In this study, we determine an upper limit
on $\epsilon_{B}$ for our X-ray sample and a measurement of
$\epsilon_{B}$ for our optical sample. We use a new method to determine an upper limit on 
$\epsilon_{B}$ with X-ray data, which relies on using 
the steep decline observed by \textit{Swift} in many X-ray
light curves. We expect that the observed flux at the
end of the steep decline is larger than the predicted flux from the
external-forward shock. Making reasonable assumptions about the other
afterglow parameters, we are able to
convert this constraint into an upper limit on $\epsilon_{B}$. 
To determine a measurement of $\epsilon_{B}$ for our
optical sample, we restrict our sample to light curves that show a
power law decline with a temporal decay $\sim 1$ at early times, $\sim 10^{2} - 10^{3}$
seconds, as expected for
external-forward shock emission. We choose this selection criteria so that the optical
emission is most likely dominated by 
external-forward shock emission. Making the same reasonable
assumptions for the other afterglow parameters and using the condition
that the observed flux from the optical light
curve is equal to the external-forward shock flux, we are able to
convert this condition into a measurement of $\epsilon_{B}$. We also
applied a consistency check for the bursts that
are in common to our X-ray and optical
samples to make sure our results for $\epsilon_{B}$ are
correct. The second goal of this study is to determine how much amplification, in
addition to shock compression, is needed to explain the results for
the $\epsilon_{B}$ upper limits$/$measurements. To quantify how much
amplification beyond shock compression is required by the
observations, we also express the results we found for the $\epsilon_{B}$
upper limits (measurements) for our X-ray (optical) sample in terms of an
amplification factor. 

This paper is organized as follows.  We begin in Section
\ref{section_2} by presenting a review of the values previous studies have found for the
microphysical afterglow parameters $\epsilon_e$ (the fraction of
energy in electrons in the shocked plasma) and 
$\epsilon_B$.  In Section \ref{section_3} (Section
\ref{optical_section}), we present 
the method we use to determine 
an upper limit (measurement) on $\epsilon_B$ and apply it to
our X-ray (optical) sample of GRBs. In Section \ref{both_x-ray_and_optical}, we use the
GRBs that are in common to both samples to perform a consistency check. We search for a possible
correlation between the kinetic energy of the blastwave and $\epsilon_B$ in 
Section \ref{bursts_with_density_and_E_vs_epsilon_B_correlation}.  
In Section \ref{AF_X-ray}, we write our results for $\epsilon_B$ for
our X-ray and optical samples in
terms of an amplification factor, which quantifies how much 
amplification -- beyond shock compression -- is required by the
observations. Lastly, in
Section \ref{conclusions_and_implications}, we discuss our results and
present our conclusions.  The convention we use for the specific flux 
$f_{\nu}$, the flux per unit frequency $\nu$, is $f_{\nu} \propto \nu^{- \beta} t^{- \alpha}$. In
this convention, $\beta$ is the spectral decay index and $\alpha$ is the
temporal decay  index. For a GRB at a given redshift $z$, when calculating
the luminosity distance to the GRB, $d_{L}$, we used the Cosmological
parameters $H_{0} = 71 \mbox{ km}/\mbox{sec}/\mbox{Mpc}$,
$\Omega_{m} = 0.27$, and $\Omega_{\Lambda} = 0.73$.

\section{Literature Review Of Values Of $\epsilon_{e}$ and $\epsilon_{B}$}\label{section_2}

The flux observed from the external-forward shock depends on 6 parameters. These
parameters are $E$, $n$, $s$, $\epsilon_{e}$, $\epsilon_{B}$, and $p$. $E$ is the isotropic
equivalent kinetic energy of the jet and $n$ is the number
density of the surrounding medium. The density is taken to be
spherically symmetric and to decrease with $r$ as $n(r) \propto r^{-s}$, where $s$ is a
constant determining the density profile of the surrounding
medium and $r$ is the distance from the center of the explosion. Two
cases are usually considered for the density profile: $s=0$ and $s=2$, which respectively correspond
to a constant density medium and a wind medium. The
microphysical parameters are $\epsilon_{e}$ and $\epsilon_{B}$, where
$\epsilon_{e}$ ($\epsilon_{B}$) is the fraction of energy in the
electrons (magnetic field) in the shocked fluid.
The microphysical parameters are taken to be constant throughout the
afterglow emission. A power law distribution of electrons, $dN_{e}/d\gamma_{e} \propto \gamma_{e} ^{-p}$
with $\gamma_{e} \ge \gamma_{i}$, where $\gamma_{e}$ is the 
Lorentz factor of the electrons, $\gamma_{i}$ is the minimum Lorentz
factor of the electrons, and $N_{e}$ is the number of electrons, is
assumed to be produced when the external-forward shock interacts with the surrounding medium. 
The power-law index of the electron distribution, $p$, is a constant known as the electron index.

In practice, it is very difficult to determine the values of the
6 afterglow parameters. The value of
$p$ and the density profile of the surrounding medium (whether we have a $s=0$ or
$s=2$ medium) can be determined from observations of the afterglow
spectral decay and temporal decay of the light curve with the
so-called ``closure'' relations \citep[e.g.][]{zhang_et_al_2006}. The remaining
4 afterglow parameters are more difficult to
determine. What is needed to determine these 4 parameters 
is observations of the afterglow emission in the 4 different spectral regimes of the synchrotron afterglow 
spectrum (we will discuss the afterglow synchrotron spectrum in more
detail in Section \ref{spectrum_at_the_end_of_the_steep_decline}). In practice,
most GRBs do not have this wealth of observations. In order to determine these 4 parameters, previous
works have either focused only on determining the afterglow parameters for bursts with high quality
data, spanning all portions of the synchrotron spectrum, or have applied
various simplifying assumptions. 

We performed a literature search for papers 
that determine values for $\epsilon_{e}$ and $\epsilon_{B}$ to get an idea
of what typical values previous works have
found. Different authors applied different techniques for finding $\epsilon_{e}$ and $\epsilon_{B}$.
When displaying the results from the literature, we did not discriminate
against any method and simply plotted every value we found. However, we
  did not consider works that made simplifying assumptions
when determining $\epsilon_{e}$ and $\epsilon_{B}$, such as
equipartition of proton and electron energy ($\epsilon_{e}$). The GRBs
for which we found $\epsilon_{e}$ and $\epsilon_{B}$ values are shown
in Table \ref{epsilon_e_and_epsilon_B_table} in Appendix
\ref{literature_values_appendix}.  Except for GRB 080928, all the GRBs in our sample have radio, optical,
and X-ray observations, allowing for a determination of all the
afterglow parameters. We included GRB 080928 in our sample because $\epsilon_{e}$ and
$\epsilon_{B}$ were able to be uniquely determined from optical and
X-ray observations \citep{rossi_et_al_2011}.

\begin{figure*}
\begin{center}$
\begin{array}{c}
\begin{array}{cc}
\hspace{-10mm}
\includegraphics[scale=0.6]{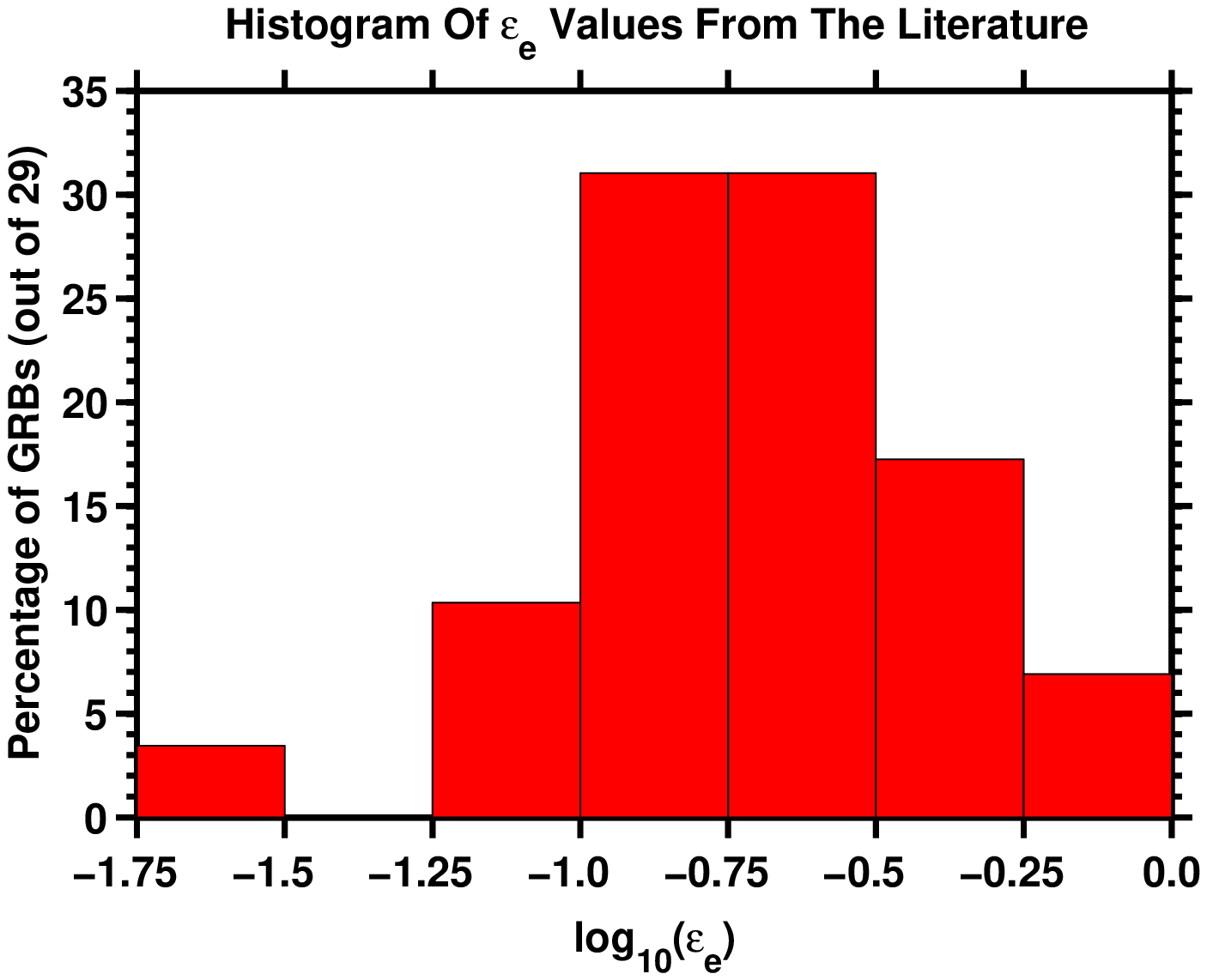} &
\hspace{0mm}
\includegraphics[scale=0.6]{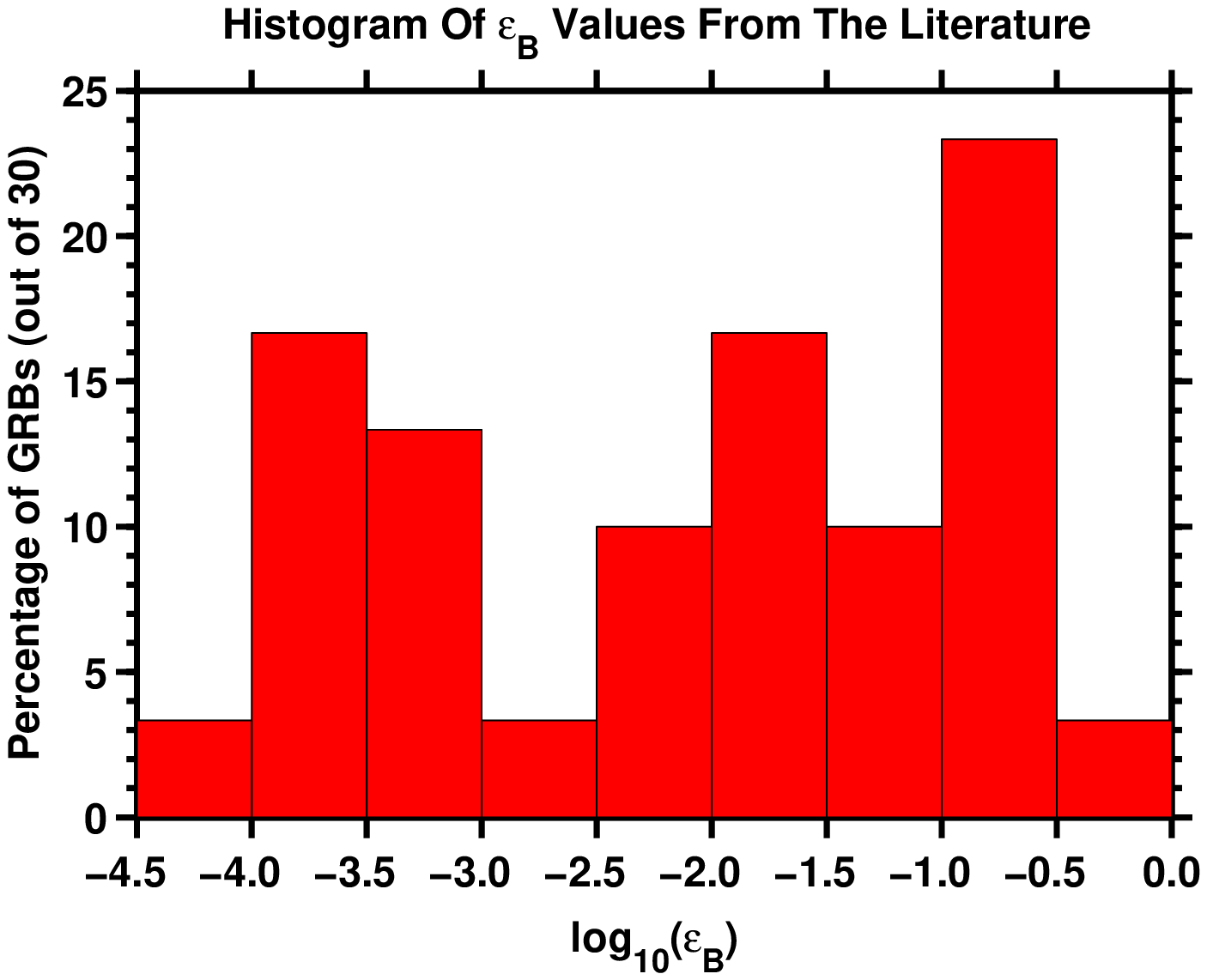}
\end{array}
\end{array}$
\end{center}
\caption{\textit{Left Panel:} Histogram of the distribution of $\epsilon_{e}$ values
we found in the literature (Table \ref{epsilon_e_and_epsilon_B_table}
in Appendix \ref{literature_values_appendix}). \textit{Right Panel:} 
Histogram of the distribution of $\epsilon_{B}$ values
we found in the literature (Table \ref{epsilon_e_and_epsilon_B_table} in  
Appendix \ref{literature_values_appendix}) \label{literature_compilation_histograms}}
\end{figure*}

The $\epsilon_{e}$ values we found in the literature for 29 GRBs are
shown in the histogram in the left panel of Figure
\ref{literature_compilation_histograms}.
There is a narrow distribution for $\epsilon_{e}$; it only varies over 
one order of magnitude, from $\sim 0.02 - 0.6$, with very few GRBs reported
to have $\epsilon_{e} < 0.1$. The mean of this distribution is 0.24 and the median is 0.22.
About 62\% of the GRBs in this sample have $\epsilon_{e} \sim 0.1-0.3$. These results for $\epsilon_{e}$ are also supported by recent simulations 
of relativistic magnetized collisionless electron-ion shocks presented in
\citealt{sironi_and_spitkovsky_2011}, where they found $\epsilon_{e} \sim 0.2$. 
The narrow distribution of $\epsilon_{e}$ values from the literature
and the results from recent simulations 
both show that $\epsilon_{e}$ does not change by much from GRB to GRB.

The $\epsilon_{B}$ values we found in the literature for 30 GRBs are
show in the histogram in the right panel of
Figure \ref{literature_compilation_histograms}. Comparing the two histograms in
Figure \ref{literature_compilation_histograms}, we can immediately see that
there is a much wider range in the distribution of $\epsilon_{B}$, with $\epsilon_{B}$
ranging from $\sim 3.5 \times 10^{-5} -0.33$. A noticeable peak, containing
about 24\% of the bursts, is seen in the bin with 
$-1 < \mbox{log}_{10} (\epsilon_{B}) < -0.5$. Two other peaks, each
containing about 17\% of the GRBs, are seen
in the bins with $-2 < \mbox{log}_{10} (\epsilon_{B}) < -1.5$ and 
$-4 < \mbox{log}_{10} (\epsilon_{B}) < -3.5$. The mean of this distribution is
$6.3 \times 10^{-2}$ and the median is 
$1.0 \times 10^{-2}$. The important point of the $\epsilon_{B}$
histogram is that $\epsilon_{B}$ varies over 4 orders of magnitude, showing that
$\epsilon_{B}$ has a wide distribution and is an uncertain parameter.

\section{Upper Limit On $\epsilon_{B}$ With \textit{Swift} X-ray Light
  Curves}\label{section_3}

\subsection{Constraining $\epsilon_{B}$ With The X-ray Light Curve Steep
Decline}\label{steep_decline_overview}

One interesting property found by \textit{Swift}
\citep{gehrels_el_al_2004} is that at early times, about 50\% of the light curves detected by
the XRT \citep[X-ray Telescope,][]{burrows_et_al_2005} display a very
rapid decline in flux, known as the steep decline \citep{gehrels_et_al_2009}. The flux during 
the steep decline typically decays as $t^{-3}$ and it
usually lasts $\sim 10^{2}-10^{3}$ sec. By extrapolating the 
BAT \citep[Burst Alert Telescope,][]{barthelmy_et_al_2005} emission
to the X-ray band, \citealt{obrien_et_al_2006a} showed that there is a continuous
transition between the end of the prompt emission and the start of the steep decline phase. This important conclusion 
lead to the interpretation that the X-ray steep decline has an origin associated 
with the end of the prompt emission.

The favored interpretation for the origin of the steep decline 
is high latitude emission \citep{kumar_and_panaitescu_2000}. Although high latitude emission
is able to explain most of the steep decline observations, some
GRBs display spectral evolution during the steep decline \citep{zhang_bin-bib_et_al_2007},
which is not expected. In any case, the steep decline cannot be produced by the external-forward shock.
Therefore, the observed flux during the steep decline should be larger than or equal to 
the flux produced by the external-forward shock.  We do however assume that the time
at which the steep decline typically ends, at about $10^{2}-10^{3}$ 
seconds, is past the deceleration time\footnote{The deceleration time marks the time when about
half of the kinetic energy of the blastwave has been transferred to
the surrounding medium.}. For our upper limit on $\epsilon_{B}$
with X-ray data, we will use the expression for the flux from the external-forward shock, 
which uses the kinetic energy of the
blastwave given by the Blandford and McKee solution \citep{blandford_and_mckee_1976}. Since
this solution is only valid for a decelerating blastwave, we need to be past the deceleration 
time for it to be applicable.

Theoretically, it is expected that that the deceleration time occurs before the end 
of of the steep decline. Depending on the density profile of the
surrounding medium, the deceleration time $t_{\mathrm{dec}}$ is given by
\begin{equation}
t_{\mathrm{dec}} = \left\{ 
\begin{array}{ll}
\hskip -7pt (220 \mbox{ sec}) E_{53} ^{1/3} n_{0} ^{-1/3} \Gamma_{2} ^{-8/3} (1+z) 
 & s=0 \\  & \\
\hskip -7pt (67 \mbox{ sec})  E_{53} A_{*,-1} ^{-1} \Gamma_{2} ^{-4} (1+z) 
& s=2 
\end{array} 
\right . \label{dec_time_eq}
\end{equation}
\citep[e.g.][]{panaitescu_and_kumar_2000}. In these expressions,
$\Gamma$ is the Lorentz factor of the shocked fluid, $z$
is the redshift, and we have adopted the usual notation 
$Q_{n} \equiv Q/10^{n}$. For $s=2$, the proportionality 
constant of the density, $A$, is normalized to the typical mass loss rate and 
stellar wind velocity of a Wolf-Rayet star, which is denoted by
$A_{*}$ and is defined as $A_{*} \equiv A/(5 \times 10^{11} \mbox{ g} \mbox{ cm}^{-1})$
\citep{chevalier_and_li_2000}. For typical GRB 
afterglow parameters of $E_{53} = 1$, $n_{0} =1$ 
(or $A_{*}=0.1$ for $s=2$), $\Gamma_{2} = 3$, and $z=2.5$, the 
deceleration time is under $100$ seconds for both $s=0$ and $s = 2$. 
Although there can be a large uncertainty in the afterglow
  parameters $E$ and $n$, $\Gamma$ is the most important parameter
  when calculating $t_{\mathrm{dec}}$ since $t_{\mathrm{dec}}$ has a
  very strong dependence on $\Gamma$. For $s=0$ ($s=2$), even if we
  take extreme parameters for $E$ and $n$, such as a high 
kinetic energy of $E_{53} = 100$ and a low density of 
$n=10^{-3} \mbox{ cm}^{-3}$ ($A_{*} = 10^{-2}$), with a
typical $\Gamma_{2} = 3-4$
  \citep[e.g.][]{molinari_et_al_2007,xue_et_al_2009,liang_et_al_2010},
$t_{\mathrm{dec}}$ is still a few hundred seconds. Thus, since the deceleration time
is less than the typical time at which the steep decline ends, the onset of the
external-forward shock emission occurs before the end of the steep decline. 

Observationally, the deceleration time is also seen to occur before the end
of the steep decline for many GRBs. If the dominant contribution to
the light curves at early times is the external-forward shock,
the light curve is expected to rise as a power law, reach a peak, 
and then decline as a power law, with the
peak signifying the deceleration time. In \citealt{liang_et_al_2010}, a
sample of optical light curves that display this peak 
is studied. In their Figure 1, for each GRB, they display both the optical
light curve and the X-ray light curve. 
For all their bursts that display a steep decline in the
X-ray light curve (except for GRB 080303 and GRB 081203A), it can be seen that the peak of the optical light
curve occurs before the end of the X-ray steep decline. We did not include
GRB 080330 and GRB 081203A in our samples. 

Since an increase in
$\epsilon_{B}$ increases $f_{\mathrm{ES}}$ (external-forward shock flux), the condition
that the X-ray flux during the steep decline should be larger than or
equal to $f_{\mathrm{ES}}$ gives an upper limit on $\epsilon_{B}$.
Since our goal is to attain the most stringent upper limit on $\epsilon_{B}$, we 
take this constraint at the end of the steep decline. Explicitly, the constraint we will use to 
find an upper limit on $\epsilon_{B}$ with X-ray data is
\begin{equation}
f_{\mathrm{EoSD}} \ge f_{\mathrm{ES}} (E , n , s , \epsilon_{e} , \epsilon_{B} , p ) . 
\label{first_x-ray_constraint}
\end{equation}
In this inequality, $f_{\mathrm{EoSD}}$ represents the observed flux at the
end of the steep decline (EoSD). We have also explicitly shown the
dependence of $f_{\mathrm{ES}}$ on the afterglow parameters. We will
now discuss the assumptions we make on the other
afterglow parameters, which will allow us to calculate an upper limit on $\epsilon_{B}$.
\subsection{The Other Afterglow Parameters}\label{X-ray_afterglow_parameter_assumptions}

\subsubsection{$E$ and $\epsilon_{e}$}\label{E_vs_epsilon_e_arguments}
Although we do not know the isotropic equivalent kinetic energy of the blastwave $E$, 
we can calculate the isotropic energy released in gamma-rays during the prompt
emission:
\begin{equation}
E_{iso} ^{\gamma} = \frac{ \mbox{fluence} \times 4 \pi d_{L}^{2}}{1+z} .
\label{gamma_ray_iso_equiv_energy_eq}
\end{equation}
In this equation, the fluence has units of ergs/$\mbox{cm}^{2}$ and
it represents the flux detected in gamma-rays, integrated
over the duration of the prompt emission. $d_{L}$ is the luminosity
distance. Since we are interested in
the fluence radiated in gamma-rays, we use the fluence detected
in the BAT band, ranging from 15-150 keV \citep{barthelmy_et_al_2005}. The fluences detected by
BAT for each GRB can be found in NASA's \textit{Swift} GRB Table and Lookup
website\footnote{
\href{http://heasarc.gsfc.nasa.gov/docs/swift/archive/grb_table/}
{http://heasarc.gsfc.nasa.gov/docs/swift/archive/grb\_table/ } }.

To convert $E_{iso} ^{\gamma}$ to $E$,
we need to know the efficiency in the conversion of kinetic energy of
the jet to prompt gamma-ray emission. Recent studies on the prompt emission
efficiency, using X-ray light curves with plateaus detected by \textit{Swift}, were
presented in \citet{granot_et_al_2006} and \citet{zhang_et_al_2007}. For the 23 GRBs for which
\citet{zhang_et_al_2007} presented results for the efficiency (see
their Table 3), more
than half of them were found to have a high efficiency  $\gtrsim$ 30\% \footnote{In Table 3 of
\citet{zhang_et_al_2007}, they present two different estimates for
the efficiency. If the shallow decay of plateaus seen in X-ray light curves is due
to energy injection, the more appropriate of the two estimates for the
efficiency is denoted as $\eta_{\gamma}(t_{\mathrm{dec}})$. 
$\eta_{\gamma}(t_{\mathrm{dec}})$ represents the efficiency in gamma-ray
radiation ($\eta_{\gamma}$) calculated at the deceleration time $t_{\mathrm{dec}}$.}, 
with a few being estimated to have
an efficiency as high as 90\%. A
high efficiency of $\sim$90\% was also found in
\citet{granot_et_al_2006}. However, \citealt{fan_and_piran_2006}
argue that bursts with X-ray plateaus should
have more moderate efficiencies $\sim 10\%$. In addition,
\citet{zhang_et_al_2007} mention that the efficiencies they calculate
for some bursts have large errors due to the uncertainty in the
microphysical parameters $\epsilon_{e}$ and $\epsilon_{B}$. 
In this work, we will take a standard choice and
calculate $E$ with the expression
\begin{equation}
E = 5 E_{\mathrm{iso}} ^{\gamma} . \label{KE_assumption}
\end{equation}
From the definition of the efficiency, 
$\eta = E^{\gamma} _{\mathrm{iso}} / (E^{\gamma} _{\mathrm{iso}} + E)
$, Equation \ref{KE_assumption} corresponds to an efficiency of $\sim 20$\%. At the end of Section
\ref{X-ray_results}, we will discuss in more detail how the
uncertainty in the efficiency affects our results. For our X-ray sample (see Section \ref{X-ray Sample}), from Equation
\ref{KE_assumption}, we found values for $E$ in the range of $10^{51}-10^{54}$ ergs, 
with a typical value $\sim 10^{53}$ ergs. The average value of $E_{53}$ for our X-ray sample is
2.8 and the median is 1.6. The fluence detected in the BAT band,
$E ^{\gamma} _{\mathrm{iso, }52}$, $z$, and $d_{L28}$ for each GRB in
our X-ray sample are shown in Table \ref{x-ray_properties_table}
 \footnote{Except for GRBs 060708, 060906, 061021, 061222A,
080906, and 081230, we obtained all the redshifts from NASA's 
\textit{Swift} GRB Table and Lookup website. For the exceptions,
we obtained the redshifts from the website on GRB redshifts
maintained by J. Greiner,
\htmladdnormallink{ http://www.mpe.mpg.de/~jcg/grbgen.html }
{http://www.mpe.mpg.de/$\sim$jcg/grbgen.html}}. 

For $\epsilon_{e}$, we assumed a value of 0.2 for all of the GRBs in
our sample. This choice for $\epsilon_{e}$ is justified from the
results of $\epsilon_{e}$ with previous afterglow studies 
(Figure \ref{literature_compilation_histograms}) and with recent results from 
simulations \citep{sironi_and_spitkovsky_2011}, as discussed in Section \ref{section_2}. 

\subsubsection{Electron power-law index and density profile}\label{p_an_s_discussion_x-ray}
In afterglow studies, $s$ and $p$ can be obtained by determining which 
closure relation the observed temporal and spectral decay indices satisfy. We
cannot use this strategy to determine $p$ and $s$ for our X-ray sample
since the external-forward shock flux is below the observed steep
decline emission. Instead, we use a fixed $p$ for all GRBs in our X-ray sample. We consider a small value
of $p=2.2$, a typical value
of $p=2.4$ for \textit{Swift} GRBs \citep{curran_et_al_2010}, and a
larger value of $p=2.8$. Previous afterglow studies have
found that the majority of afterglow observations are better described by a constant
density medium
\citep[e.g.][]{panaitescu_and_kumar_2002,schulze_et_al_2011}. However,
there are still a number of cases where the wind medium is a better model for the 
afterglow observations. Therefore, we will consider both $s=0$ and
$s=2$ when displaying the results for the upper limit on $\epsilon_{B}$. 

\subsubsection{Density}

The density of the medium in the vicinity of GRBs is a highly uncertain parameter. A histogram of values of $n$
determined by previous afterglow modelling studies can be found in
Figure 9 of \citet{soderberg_et_al_2006}, which shows $n$ to vary
over 5 orders of magnitude, ranging from 
$\sim 10^{-3} \mbox{ cm}^{-3}$ to $\sim 10^{2} \mbox{ cm}^{-3}$.
In Section \ref{X-ray_results}, we will discuss in more
detail how the uncertainty in the density affects our results for the
$\epsilon_{B}$ upper limits.  

\subsection{The X-ray Sample}\label{X-ray Sample}
For our constraint on $\epsilon_{B}$ with X-ray data, we only consider X-ray
data detected by the XRT on board \textit{Swift}. We used the X-ray light curves presented in 
\citet{butler_et_al_2007}\footnote{ 
\htmladdnormallink{ http://butler.lab.asu.edu/swift/ }
{http://butler.lab.asu.edu/swift/}}. With the exception of two cases, we
only consider bursts that display a steep decline in their X-ray light curve (see below).
After the end of the steep decline, GRBs display a variety of temporal decays \citep{evans_et_al_2009}. 
Our sample can be divided into 4 different subgroups, based on
the temporal decay after the steep decline:
\begin{enumerate}
\item \underline{Steep Decline To Plateau:} In this subgroup, GRBs display a plateau
after the steep decline ($\sim$ 73\% of our sample). In
Table \ref{x-ray_properties_table}, we display the time at the end of the steep decline in
units of $10^{2}$ sec, $t_{2, \mathrm{EoSD}}$, and the observed flux at the end of the steep decline
at 1 keV, $f_{1 \mathrm{keV}, \mathrm{EoSD}}$, in $\mu$Jy. 
\item \underline{Steep Decline To Normal Decline:} In this subgroup,
  GRBs display a temporal decay of $\alpha \sim 1$ after the steep
  decline ($\sim$ 18\% of our sample). In Table \ref{x-ray_properties_table}, we
show the time and the flux at the end of the steep decline. 
\item \underline{Clear Steep Decline But Not A Clear End To The} 
 \underline {Steep Decline:} In this subgroup, it cannot be determined where the
  steep decline ends (the XRT observations end before the steep
  decline ends). The
following GRBs fall into this subgroup: 050315, 060202, 070419A, 071122, and 090516.
For these GRBs, in Table \ref{x-ray_properties_table}, we show the 
flux and the time of the last steep decline point observed. For these
GRBs, to be sure that the last steep decline point observed is past the deceleration time,
we made sure that it is at a few hundred seconds. 
\item \underline{No Clear Steep Decline Seen, Just Plateau:} Two of the GRBs in our X-ray sample,
050401 and 060927, do not display a steep decline. The first observation of the X-ray light curve for these bursts
is during the plateau. We did not
remove these GRBs from our X-ray sample because they are also part
of our optical sample (bursts with both X-ray and
optical data are important because they allow us to cross-check our
results, see Section \ref{both_x-ray_and_optical}). 
For these 2 bursts, we considered the first observation in the X-ray light curve
for our constraint so that we have the least amount of energy
injection. We made sure that this point is at least at a few hundred seconds so that we can be confident that the
onset of the external-forward shock emission has occurred. For these two bursts, we show the time and
the flux of the first X-ray observation of the plateau in Table \ref{x-ray_properties_table}.
\end{enumerate}
In addition, 25\% of the GRBs in our X-ray sample display X-ray flares
during the steep decline. We only consider bursts where the
X-ray flare ends before the end of the steep decline because it is
difficult to determine the flux and time at the end of the steep
decline for bursts that show X-ray flares near the end of the steep
decline. It is fine to consider these bursts because observationally, after the X-ray flare, the 
X-ray light curve is seen to return to the same temporal decay 
prior to the X-ray flare (e.g. \citealt{chincarini_et_al_2007}). Lastly, GRB 051221A is the
only short GRB in our X-ray sample; all the other bursts in our X-ray
sample are long GRBs. 
%
%
%

\begin{small}
\begin{longtable*}{lcccccccc}
\caption{Properties of X-ray Sample \label{x-ray_properties_table}}\\
\hline
GRB & 
$z$  &
$d_{L28}$  &
Fluence & 
$E ^{\gamma} _{\mathrm{iso, }52}$  & 
$t_{2, \mathrm{EoSD}}$  & 
$f_{1 \mathrm{keV, EoSD}}$ &
$\mbox{log}_{10} (\epsilon_{B})$ &
$\mbox{log}_{10} (\epsilon_{B})$ \\
 & 
 &
 &
$[ \times 10^{-6} \mbox{ ergs}/\mbox{cm}^{2}$] & 
 & 
 & 
 [$\mu$Jy] &
 ($s=0$) &
  ($s=2$)\\
\hline
\endfirsthead
\caption{Properties of X-ray Sample (Continued)}\\
\hline
GRB & 
$z$  &
$d_{L28}$  &
Fluence & 
$E ^{\gamma} _{\mathrm{iso, }52}$  & 
$t_{2, \mathrm{EoSD}}$  & 
$f_{1 \mathrm{keV, EoSD}}$ &
$\mbox{log}_{10} (\epsilon_{B})$ &
$\mbox{log}_{10} (\epsilon_{B})$ \\
 & 
 &
 &
$[ \times 10^{-6} \mbox{ ergs}/\mbox{cm}^{2}$] & 
 & 
 & 
 $\mu$Jy &
 $s=0$ &
 $s=2$  \\
\hline
\endhead
\hline 
\endfoot
\hline
\endlastfoot
050315	&	1.949	&	4.71	&	3.22	&	3.04	&	4	&	3	&	-5.2	&	-6.0	\\
\textbf{050401}	&	2.9	&	7.67	&	8.22	&	15.56	&	2	&	80	&	-4.7	&	-5.4	\\
\textbf{050721}	&	2.5	&	6.40	&	3.62	&	5.32	&	4	&	30	&	-4.2	&	-5.0	\\
050803	&	0.422	&	0.71	&	2.15	&	0.10	&	3	&	10	&	-3.8	&	-5.4	\\
050814	&	5.3	&	15.76	&	2.01	&	9.95	&	9	&	1	&	-5.4	&	-6.0	\\
\textbf{051221A}	&	0.547	&	0.97	&	1.15	&	0.09	&	6	&	5	&	-3.5	&	-4.9	\\
060108	&	2.03	&	4.95	&	0.37	&	0.38	&	6	&	0.8	&	-4.2	&	-5.5	\\
\textbf{060111B}	&	2	&	4.86	&	1.60	&	1.58	&	2	&	8	&	-4.6	&	-5.8	\\
060115	&	3.53	&	9.72	&	1.71	&	4.48	&	8	&	1	&	-5.2	&	-5.9	\\
\textbf{060210}	&	3.91	&	10.99	&	7.66	&	23.66	&	8	&	40	&	-4.4	&	-4.7	\\
\textbf{060418}	&	1.49	&	3.37	&	8.33	&	4.79	&	5	&	40	&	-4.3	&	-4.9	\\
060502A	&	1.51	&	3.43	&	2.31	&	1.36	&	3	&	7	&	-4.6	&	-5.7	\\
\textbf{060607A}	&	3.082	&	8.25	&	2.55	&	5.35	&	5	&	60	&	-3.6	&	-4.3	\\
060707	&	3.425	&	9.37	&	1.60	&	3.99	&	9	&	3	&	-4.5	&	-5.2	\\
060708	&	1.92	&	4.62	&	0.49	&	0.45	&	2	&	10	&	-3.7	&	-5.2	\\
060714	&	2.71	&	7.06	&	2.83	&	4.78	&	3	&	10	&	-4.8	&	-5.7	\\
060729	&	0.54	&	0.96	&	2.61	&	0.20	&	6	&	7	&	-3.8	&	-5.1	\\
060814	&	0.84	&	1.65	&	14.60	&	2.71	&	8	&	8	&	-5.0	&	-5.6	\\
060906	&	3.686	&	10.24	&	2.21	&	6.21	&	4	&	0.8	&	-5.9	&	-6.7	\\
060926	&	3.208	&	8.66	&	0.22	&	0.49	&	2	&	4	&	-3.8	&	-5.4	\\
\textbf{060927}	&	5.6	&	16.81	&	1.13	&	6.08	&	0.8	&	8	&	-5.3	&	-6.6	\\
061021	&	0.3463	&	0.56	&	2.96	&	0.09	&	3	&	10	&	-3.9	&	-5.5	\\
061110A	&	0.758	&	1.45	&	1.06	&	0.16	&	5	&	5	&	-3.6	&	-6.0	\\
\textbf{061121}	&	1.314	&	2.88	&	13.70	&	6.19	&	2	&	40	&	-5.1	&	-5.8	\\
061222A	&	2.088	&	5.13	&	7.99	&	8.55	&	2	&	40	&	-4.9	&	-5.7	\\
070110	&	2.352	&	5.94	&	1.62	&	2.14	&	4	&	8	&	-4.3	&	-5.3	\\
070306	&	1.497	&	3.39	&	5.38	&	3.12	&	7	&	4	&	-5.0	&	-5.6	\\
\textbf{070714B}	&	0.92	&	1.85	&	0.72	&	0.16	&	4	&	6	&	-3.5	&	-5.0	\\
070802	&	2.45	&	6.24	&	0.28	&	0.40	&	5	&	0.8	&	-4.2	&	-5.5	\\
071122	&	1.14	&	2.41	&	0.58	&	0.20	&	8	&	0.8	&	-4.1	&	-5.4	\\
080310	&	2.43	&	6.18	&	2.30	&	3.22	&	10	&	2	&	-4.8	&	-5.4	\\
080413A	&	2.433	&	6.19	&	3.50	&	4.91	&	2	&	10	&	-5.1	&	-6.0	\\
080430	&	0.767	&	1.47	&	1.20	&	0.19	&	2	&	10	&	-3.9	&	-5.5	\\
\textbf{080607}	&	3.036	&	8.10	&	24.00	&	49.07	&	3	&	90	&	-5.2	&	-5.5	\\
\textbf{080721}	&	2.591	&	6.68	&	12.00	&	18.75	&	0.7	&	900	&	-4.3	&	-5.1	\\
080905B	&	2.374	&	6.00	&	1.80	&	2.42	&	2	&	20	&	-4.3	&	-5.4	\\
080906	&	2.1	&	5.16	&	3.50	&	3.78	&	7	&	20	&	-4.0	&	-4.7	\\
080916A	&	0.689	&	1.29	&	4.00	&	0.50	&	3	&	10	&	-4.4	&	-5.7	\\
081007	&	0.5295	&	0.93	&	0.71	&	0.05	&	2	&	8	&	-3.5	&	-5.3	\\
\textbf{081008}	&	1.9685	&	4.77	&	4.30	&	4.14	&	6	&	20	&	-4.2	&	-4.9	\\
081230	&	2	&	4.86	&	0.82	&	0.81	&	3	&	6	&	-4.1	&	-5.3	\\
\textbf{090418A}	&	1.608	&	3.71	&	4.60	&	3.05	&	2	&	20	&	-4.8	&	-5.8	\\
090516A	&	4.109	&	11.66	&	9.00	&	30.08	&	6	&	10	&	-5.4	&	-5.7	\\
090519	&	3.85	&	10.79	&	1.20	&	3.62	&	5	&	0.8	&	-5.4	&	-6.2	\\
090529	&	2.625	&	6.79	&	0.68	&	1.09	&	20	&	0.5	&	-4.3	&	-5.1	\\
090618	&	0.54	&	0.96	&	105.00	&	7.86	&	4	&	200	&	-4.9	&	-5.3	\\
090926B	&	1.24	&	2.68	&	7.30	&	2.95	&	5	&	10	&	-4.8	&	-5.6	\\
091029	&	2.752	&	7.19	&	2.40	&	4.16	&	6	&	1	&	-5.5	&	-6.2	\\
091109A	&	3.076	&	8.25	&	1.60	&	3.35	&	5	&	2	&	-5.0	&	-5.8	\\
100302A	&	4.813	&	14.06	&	0.31	&	1.33	&	8	&	1	&	-4.2	&	-5.2	\\
100425A	&	1.755	&	4.14	&	0.47	&	0.37	&	3	&	7	&	-3.6	&	-5.0	\\
100513A	&	4.772	&	13.92	&	1.40	&	5.91	&	9	&	7	&	-4.1	&	-4.8	\\
100621A	&	0.542	&	0.96	&	21.00	&	1.58	&	4	&	20	&	-5.0	&	-5.8	\\
100704A	&	3.6	&	9.95	&	6.00	&	16.23	&	6	&	9	&	-5.1	&	-5.5	\\
100814A	&	1.44	&	3.23	&	9.00	&	4.85	&	6	&	9	&	-5.0	&	-5.6	\\
100906A	&	1.727	&	4.05	&	12.00	&	9.09	&	3	&	20	&	-5.2	&	-5.9	\\
110808A	&	1.348	&	2.98	&	0.33	&	0.16	&	3	&	3	&	-3.6	&	-5.2	\\
110818A	&	3.36	&	9.16	&	4.00	&	9.67	&	20	&	1	&	-5.3	&	-5.5	\\
111008A	&	4.9898	&	14.68	&	5.30	&	23.95	&	3	&	9	&	-5.5	&	-6.1	\\
111228A	&	0.72	&	1.36	&	8.50	&	1.15	&	5	&	8	&	-4.8	&	-5.7	\\
\hline
\caption{\footnotesize{ This table displays the properties of our
X-ray sample. The
GRBs that are in bold are also part of our optical sample. The second
and third columns show the redshift and the
luminosity distance $d_{L}$ (in units of $10^{28}$ cm), respectively. The fourth column
shows the fluence detected by BAT in units of $10^{-6} \mbox{ergs}/\mbox{cm}^{2}$. The next column
shows $E ^{\gamma} _{\mathrm{iso, }52}$, the isotropic equivalent
energy released in gamma-rays during the prompt emission, in units of $10^{52} \mbox{ ergs}$.
$t_{2, \mathrm{EoSD}}$ represents the time
at the end of the steep decline (EoSD) in
units of $10^{2}$ seconds. The column $f_{1 \mathrm{keV, EoSD}}$ shows 
the specific flux at 1 keV at the end of
the steep decline, in units of $\mu$Jy. The last two columns show the
upper limits on $\epsilon_{B}$, assuming $p=2.4$. One column shows the
results for a constant density medium ($s=0$) assuming $n = 1 \mbox{
  cm}^{-3}$ (filled-in histogram in the Top-Right panel of 
 Figure \ref{x-ray_epsilon_upper_limit_histograms}) and the other column
shows the results for a wind medium ($s=2$) assuming $A_{*} = 0.1$ 
(un-filled histogram in the Top-Right panel of Figure
\ref{x-ray_epsilon_upper_limit_histograms}).} }
\end{longtable*}
\end{small}
%
%
%
%
\subsection{Expected External-Forward Shock Emission At The End Of The Steep
  Decline}\label{spectrum_at_the_end_of_the_steep_decline}

The synchrotron afterglow spectrum consists of four power-law segments that are smoothly joined
together at three characteristic frequencies of synchrotron
emission \citep*[e.g.][]{sari_et_al_1998,granot_and_sari_2002}. These
three characteristic frequencies are: $\nu_{a}$, the synchrotron self-absorption frequency,
$\nu_{i}$ (also commonly referred to as $\nu_{m}$), 
the frequency of the photons emitted by the 
power-law distribution of injected electrons with the
smallest energy, and $\nu_{c}$, the cooling frequency corresponding to electrons cooling on a dynamical time.
For this work, we will only consider
the standard case for the ordering of the characteristic frequencies, 
the slow cooling case, where $\nu_{a} < \nu_{i} < \nu_{c}$. 
One argument against the fast cooling case ($\nu_{a} < \nu_{c} < \nu_{i}$) is that if the observing frequency is between
$\nu_{c}$ and $\nu_{i}$, the spectrum should be $f_{\nu} \propto
\nu^{-1/2}$; however, the spectral index $\beta = 0.5$
disagrees with the typical observed afterglow spectral index $\beta \approx 0.9$ \citep*[e.g.][]{piro_2001}.
The next question we need to consider is where the X-ray band lies at
the end of the steep decline (here, we consider 1 keV for the X-ray
band because the light curves we used are plotted at this
energy). The two possibilities for the spectral regime of the X-ray
band are $\nu_{i} < 1 \mbox{ keV} <\nu_{c}$ or $\nu_{c} < 1 \mbox{
  keV}$. We rule out $\nu_{c} < 1 \mbox{ keV}$ with the following two arguments. 

First, we compare the observed flux at the end of the
steep decline, $f_{\mathrm{1 keV, EoSD}}$, to the flux predicted by the 
external-forward shock at the same time, if $\nu_{c} < 1 \mbox{ keV}$
(defined as $f_{\mathrm{pred}}$). For $s=0$ , $f_{\mathrm{pred}}$ is given by \citep{granot_and_sari_2002}
\begin{eqnarray}
&& f_{\mathrm{pred}} = 0.855 (p-0.98) 10^{\frac{3p+2}{4}} 8.64^{\frac{3p-2}{4}}
e^{1.95p}  \nonumber \\ 
&& \times (1+z)^{\frac{p+2}{4}}  (\bar{\epsilon}_{e,-1}) ^{p-1} \epsilon_{B}
^{\frac{p-2}{4}} E_{53} ^{\frac{p+2}{4}} (t_{\mathrm{2, EoSD}}) ^{-\frac{(3p-2)}{4}}  \nonumber \\
&& \times d_{L28} ^{-2} \nu_{14} ^{-\frac{p}{2}} \mbox{ mJy ,} \label{ES_eq_1keV_greater>nu_c}
\end{eqnarray}
where $\bar{\epsilon}_{e,-1} \equiv (p-2)/(p-1) \epsilon_{e,-1}$. For 
$\nu_{c} < 1 \mbox{ keV}$, the external-forward shock flux is
independent of the density and the $s=2$ expression is almost identical. 
When calculating $f_{\mathrm{pred}}$, for each of the bursts in our X-ray
sample, we assumed a standard $p = 2.4$, $\epsilon_{e} = 0.2$, and 
$\nu_{14} = 2.4 \times 10^{3}$, the frequency corresponding to 1 keV. 
For the parameters $E$, $t$, $z$, and $d_{L}$, we used the
values given in Table \ref{x-ray_properties_table} for each burst 
($E = 5 E_{iso} ^{\gamma}$). The remaining parameter we need to compute $f_{\mathrm{pred}}$
is $\epsilon_{B}$. Since $\epsilon_{B}$ is raised to the power of $(p-2)/4$, for a typical $p \sim 2-3$,
the dependence on $\epsilon_{B}$ is very weak. When calculating $f_{\mathrm{pred}}$, 
we assumed a low value of $\epsilon_{B} = 10^{-3}$. 

We computed the ratio $f_{\mathrm{pred}}/f_{\mathrm{1 keV, EoSD}}$ for
all the GRBs in our X-ray sample and found that  
$f_{\mathrm{pred}}/f_{\mathrm{1 keV, EoSD}} > 1$ for all the bursts and
$f_{\mathrm{pred}}/f_{\mathrm{1 keV, EoSD}} > 10$ for $54/60$ bursts.
The mean value of $f_{\mathrm{pred}}/f_{\mathrm{1 keV, EoSD}}$ is 50
and the median value is 34.
This means that the predicted flux from the external-forward shock, when $\nu_{c} < \mbox{1 keV}$,
over-predicts the observed flux at the end of the steep
decline by a factor that is larger than 10 for the majority of the
bursts. Therefore, the assumption that $\nu_c < 1$ keV is incorrect.
This is a robust conclusion because the X-ray flux from the
external-forward shock, when $\nu_c < 1$ keV, basically only depends 
on $\sim \epsilon_e E$ (see Equation \ref{ES_eq_1keV_greater>nu_c}),
which cannot be decreased by a factor of $>10$ without introducing 
serious efficiency problems in producing the prompt gamma-rays.  Even if we allow for an 
uncertainty of a factor of $\sim 2-3$ in both $\epsilon_e$ and $E$, this is not enough to decrease
$f_{\mathrm{pred}}/f_{\mathrm{1 keV, EoSD}}$ below 1 for the majority of bursts in our X-ray sample. 

Before continuing, we want to add that $f_{\mathrm{pred}}$ (Equation
\ref{ES_eq_1keV_greater>nu_c}) also has a dependence
on the Compton-$Y$ parameter: $f_{\mathrm{pred}} \propto (1 + Y)^{-1}$. With this
dependence, if the Compton-$Y$ parameter is large, then it is
possible for $f_{\mathrm{pred}}$ to decrease below 
$f_{\mathrm{1 keV, EoSD}}$. For a few bursts in our X-ray sample,
we performed a detailed numerical calculation of the external-forward shock flux
with the formalism presented in \citet{barniol_duran_and_kumar_2011},
which includes a detailed treatment of Compton-$Y$ with Klein-Nishina
effects. From this calculation, we also found that $f_{\mathrm{pred}}$ over-predicts
$f_{\mathrm{1 keV, EoSD}}$ by a factor larger than 10, which means
that the Inverse Compton (IC) cooling of electrons producing 1 keV synchrotron photons 
is a weak effect (even when $\epsilon_B$ is small). Numerically, we also found that without making 
any assumption about the location of $\nu_c$, solutions to the
constraint  $ f_{\mathrm{ES, 1 keV}}\leq f_{1keV,EoSD}$ were only
found when 1 keV $<\nu_c$ 
(when $\epsilon_B$ is small such 
that 1 keV $< \nu_c$, it turns out that IC cooling of electrons producing 1 keV photons takes place 
in the Klein-Nishina regime at these early times of $\sim$ few $\times 100$ sec).

Another argument in favor of the spectral regime of the X-ray band 
being $\nu_{i} < \mbox{1 keV} <  \nu_{c}$ at
the end of the steep decline comes from the extrapolation of $\nu_{c}$ at late times to
the end of the steep decline. In \citet{liang_et_al_2008}, they made fits
to late time XRT light curves during the normal decline phase and they also
provided the value of $\nu_{c}$
during the geometrical midpoint of the normal decline 
phase\footnote{The geometrical midpoint of the normal decline is defined by 
$\mbox{log}_{10} t = (\mbox{log}_{10} t_{1} + \mbox{log}_{10} t_{2})/2$
, where $t_{1}$ represents the time of the first observation of the normal
decline and $t_{2}$ represents the time of the last observation of the normal
decline.}. For this argument, we focus on GRBs that are in common to our
X-ray sample and to the sample of \citet{liang_et_al_2008}. For these
bursts, we extrapolate $\nu_{c}$ at late times to the end of the steep
decline. In \citet{liang_et_al_2008}, they
only considered a constant density medium, where $\nu_{c} \propto t^{-1/2}$. The results of the extrapolation of $\nu_{c}$ are shown in Table \ref{extrapolation_table}. 
\begin{table}[t]
\begin{small}
\begin{tabular}{c c c c c}
\hline
GRB & $t_{\mathrm{late},4}$ & $t_{\mathrm{EoSD},2}$ &
$\nu_{\mathrm{c,late}}$ (keV) & $\nu_{\mathrm{c, EoSD}}$ (keV)  \\
\hline
050315	&	$	3.5 	$	&	4	&	0.17	&	1.6	\\
050401	&	$	6.9 	$	&	2	&	4.1	&	77	\\
051221A	&	$	1.7 	$	&	6	&	4.1	&	22	\\
060210	&	$	3.2	$	&	8	&	0.97	&	6.1	\\
060502A	&	$	0.42	$	&	3	&	4.1	&	15	\\
060714	&	$	3.5	$	&	3	&	3.6	&	39	\\
060729	&	$	40 	$	&	6	&	1.4	&	36	\\
060814	&	$	2.0 	$	&	8	&	4.1	&	21	\\
060926	&	$	0.26 $	&	2	&	6.7	&	24	\\
061121	&	$	1.1 	$	&	2	&	4.2	&	31	\\
070110	&	$	14 	$	&	4	&	4	&	74	\\
\hline
\end{tabular}
\end{small}
\caption{\footnotesize{This table shows the extrapolation of $\nu_{c}$
from late times to the end of the steep decline for GRBs in common to 
our sample and to the sample in \citet{liang_et_al_2008} (first column). The second column shows
$t_{\mathrm{late},4}$, the late time in units of $10^{4}$ seconds at which \citet{liang_et_al_2008}
determined $\nu_{c}$. The third column shows $t_{\mathrm{EoSD},2}$,
the time at the end of the steep decline in units of $10^{2}$
seconds. $\nu_{\mathrm{c, late}}$, given in keV, is the value found in
\citet{liang_et_al_2008} for $\nu_{c}$ at
$t_{\mathrm{late}}$. The last column shows $\nu_{\mathrm{c, EoSD}}$ in
keV. $\nu_{\mathrm{c, EoSD}}$ is found by extrapolating
$\nu_{\mathrm{c, late}}$ to $t_{\mathrm{EoSD}}$. Since
\citet{liang_et_al_2008} assume a constant density medium, we take a
constant density medium for all the GRBs in this sample when
making the extrapolation of $\nu_{c}$ to $t_{\mathrm{EoSD}}$. \label{extrapolation_table}}}
\end{table}
In Table \ref{extrapolation_table}, we find that at the end of the steep decline, 
1 keV $< \nu_{c, \mathrm{ EoSD}}$ for all GRBs. This
further confirms our choice that $\nu_{i} < \mbox{ 1 keV} < \nu_{c}$
at the end of the steep decline\footnote{Including energy injection will make the values of 
$\nu_{\mathrm{c, EoSD}}$ in Table \ref{extrapolation_table} larger,
making the conclusion that $\nu_{i} < \mbox{ 1 keV} < \nu_{c}$ 
at the end of the steep decline more robust. For the bursts that 
have plateaus in their X-ray light curve, energy injection needs to be considered. During the energy
injection episode, $E$ increases as $E \propto t^{1-q}$ \citep{zhang_et_al_2006}, where $q$ is
a positive constant that satisfies $0 \le q \le 1$. Therefore, between the
end of the plateau and the end of the steep decline, since $\nu_{c}
\propto E^{-1/2} t^{-1/2}$, $\nu_{c} \propto t^{-1 + (q/2)}$. This
time evolution of $\nu_{c}$ is steeper than $\nu_{c} \propto t^{-1/2}$ 
without energy injection for $s=0$ (\citet{liang_et_al_2008} only
considered $s=0)$.}.

The knowledge of the spectral regime regime at the end of the steep
decline allows us to write in an explicit expression for
$f_{\mathrm{ES}}$ in Equation \ref{first_x-ray_constraint}:
\begin{equation}
\frac{f_{\mathrm{1 keV, EoSD}}}{\mbox{mJy}} \ge 
\left \{\begin{array}{ll}
\hskip -7pt 0.461(p-0.04) 10^{\frac{3p+1}{4}} 8.64^{\frac{3(p-1)}{4}}
e^{2.53p}  &  \\ 
 \times (1+z)^{\frac{p+3}{4}} (\bar{\epsilon}_{e,-1})^{p-1}
\epsilon_B^{\frac{p+1}{4}} n_0^{\frac{1}{2}} E_{53}^{\frac{p+3}{4}} &  \\ 
 \times (t_{\rm 2, EoSD}) ^{-\frac{3(p-1)}{4}}
d_{L28}^{-2}\nu_{14}^{-\frac{(p-1)}{2}}  \mbox{ $s=0$} \\  & \\
\hskip -7pt 3.82(p-0.18) 10^{\frac{3p-1}{4}} 8.64^{\frac{3p-1}{4}} 
e^{2.54p}  &  \\ 
\times (1+z)^{\frac{p+5}{4}} (\bar{\epsilon}_{e,-1}) ^{p-1} 
\epsilon_B^{\frac{p+1}{4}} A_{*,-1} E_{53}^{(p+1)/4}  & \\ 
\times (t_{\rm 2, EoSD}) ^{-\frac{(3p-1)}{4}}
d_{L28}^{-2}\nu_{14}^{-\frac{(p-1)}{2}}  \mbox{ $s=2$} .
\end{array} 
\right. \label{x-ray_constraint}  
\end{equation}
On the left hand side of this inequality, we have the observed X-ray flux at the end of the steep decline and
on the right hand side we have the expression for the
external-forward shock flux when $\nu_{i} < \mbox{ 1 keV} < \nu_{c}$
\citep{granot_and_sari_2002}. The notation used for
$\bar{\epsilon}_{e}$ and $A_{*}$ is defined as 
$\bar{\epsilon}_{e,-1} \equiv \bar{\epsilon}_{e}/10^{-1}$ and $A_{*,-1}
\equiv A_{*}/10^{-1}$. The expressions in
Equation \ref{x-ray_constraint} are only valid for $p>2$, which we
consider in this work (for $p<2$, see \citealt{bhattacharya_2001} and \citealt{resri_and_bhattacharya_2008}).

Before displaying our results for the $\epsilon_{B}$ upper limit for our entire X-ray sample,
we show a simple calculation to get an idea of what values to expect
for the $\epsilon_{B}$ upper limits from the X-ray constraint given in Equation
\ref{x-ray_constraint}. For a standard $p = 2.4$, the X-ray constraint is
\begin{equation}
\frac{f_{\mathrm{1 keV, EoSD}}}{\mbox{mJy}} \ge 
\left \{\begin{array}{ll}
\hskip -7pt 5.0 \times 10^{1} t_{2, \mathrm{EoSD}} ^{-1.05} \epsilon_{e,-1} ^{1.4} E_{53} ^{1.35}
n_{0} ^{0.5} \epsilon_{B} ^{0.85}  & s=0 \\ & \\
\hskip -7pt 7.0 \times 10^{2} t_{2, \mathrm{EoSD}} ^{-1.55} \epsilon_{e,-1} ^{1.4} E_{53} ^{0.85}
A_{*,-1} \epsilon_{B} ^{0.85}  & s=2 .
\end{array} 
\right. \label{simplified_x-ray_constraint}
\end{equation}
For this calculation, we used the average $z=2.5$ for
\textit{Swift} GRBs \citep{gehrels_et_al_2009} (with a corresponding
$d_{L28}=6.4$) and $\nu_{14}$ corresponding to 1 keV. Solving for $\epsilon_{B}$, the
upper limit depends on the afterglow parameters 
as\footnote{For bursts with plateaus in their X-ray light
    curve, it is possible that energy injection begins before the
    steep decline ends. However, even with energy injection, there
    still exists a self-similar solution for the energy (Equation 52
    of \citealt{blandford_and_mckee_1976}). For both $s=0$ and $s=2$, we calculated the
    external-forward shock synchrotron flux with this new self-similar
  solution and then calculated the upper limit on $\epsilon_{B}$ as
  shown in Equation \ref{epsilon_B_afterglow_parameters}. We found
  that the $\epsilon_{B}$ upper limits are affected by less than a
  factor of $\sim 2$. Thus, even if energy injection begins before the end of
  the steep decline, it has very little to no effect on our
  $\epsilon_{B}$ upper limit results.}  
\begin{equation}
\epsilon_{B} \le
\left \{\begin{array}{ll}
\hskip -7pt 1.0 \times 10^{-2} \left ( \frac{f_{1 keV, \mathrm{EoSD}}}{\mbox{mJy}} \right )^{1.18}
t_{2, \mathrm{EoSD}} ^{1.24} & \\ 
\times \epsilon_{e,-1} ^{-1.65} E_{53} ^{-1.59} n_{0} ^{-0.59}  & s=0 \\ & \\
\hskip -7pt 4.5 \times 10^{-4} \left ( \frac{f_{1 keV, \mathrm{EoSD}}}{\mbox{mJy}} \right )^{1.18}
t_{2, \mathrm{EoSD}} ^{1.82} & \\
\times \epsilon_{e,-1} ^{-1.65} E_{53} ^{-1} A_{*,-1} ^{-1.18}  & s=2 .
\end{array} 
\right. \label{epsilon_B_afterglow_parameters}
\end{equation}
The median values for our X-ray sample
for the parameters $f_{\mathrm{1 keV, EoSD}}$, $t_{2, \mathrm{EoSD}}$,
and $E_{53}$ are $8 \times 10^{-3}$ mJy, 4, and 1.6, respectively. 
Using these median values and $\epsilon_{e,-1} = 2$, 
the upper limit on $\epsilon_{B}$ becomes 
\begin{equation}
\epsilon_{B} \le 
\left \{\begin{array}{ll}
\hskip -7pt 2.8 \times 10^{-5} \times n_{0} ^{-0.59} & s=0 \\ & \\
\hskip -7pt 3.7 \times 10^{-6} \times A_{*-1} ^{-1.18} & s=2 .
\label{getting_upper_limit_idea}
\end{array} 
\right. 
\end{equation}
For a standard $n_{0} = 1$ and $A_{*,-1} = 1$, it can be
seen that the $\epsilon_{B}$ upper limit is lower for $s=2$. This is expected
because for $A_{*,-1} = 1$, there is a larger density for the
surrounding medium within a typical deceleration radius of $10^{17}
\mbox{ cm}$.

In Equation \ref{getting_upper_limit_idea}, the explicit dependence of
the $\epsilon_{B}$ upper limit on the density is shown for $p=2.4$. In
the next subsection, we will display the results of the $\epsilon_{B}$
upper limits for our entire X-ray sample. To keep the
density-dependence, we will display histograms of upper limits on the quantity 
$\epsilon_{B} n_{0} ^{0.59}$ ($\epsilon_{B} A_{*-1} ^{1.18}$) for
$s=0$ ($s=2$) for $p=2.4$, or $\epsilon_{B} n_{0} ^{2/(p+1)}$ 
($\epsilon_{B} A_{*-1} ^{4/(p+1)}$) for $s=0$ ($s=2$) for a general
$p$ (see Equation \ref{x-ray_constraint}).

\subsection{$\epsilon_{B}$ Upper Limits For Our X-ray
Sample}\label{X-ray_results} 

\begin{figure*}
\begin{center}$
\begin{array}{c}
\begin{array}{cc}
\hspace{-10mm}
\includegraphics[scale=0.6]{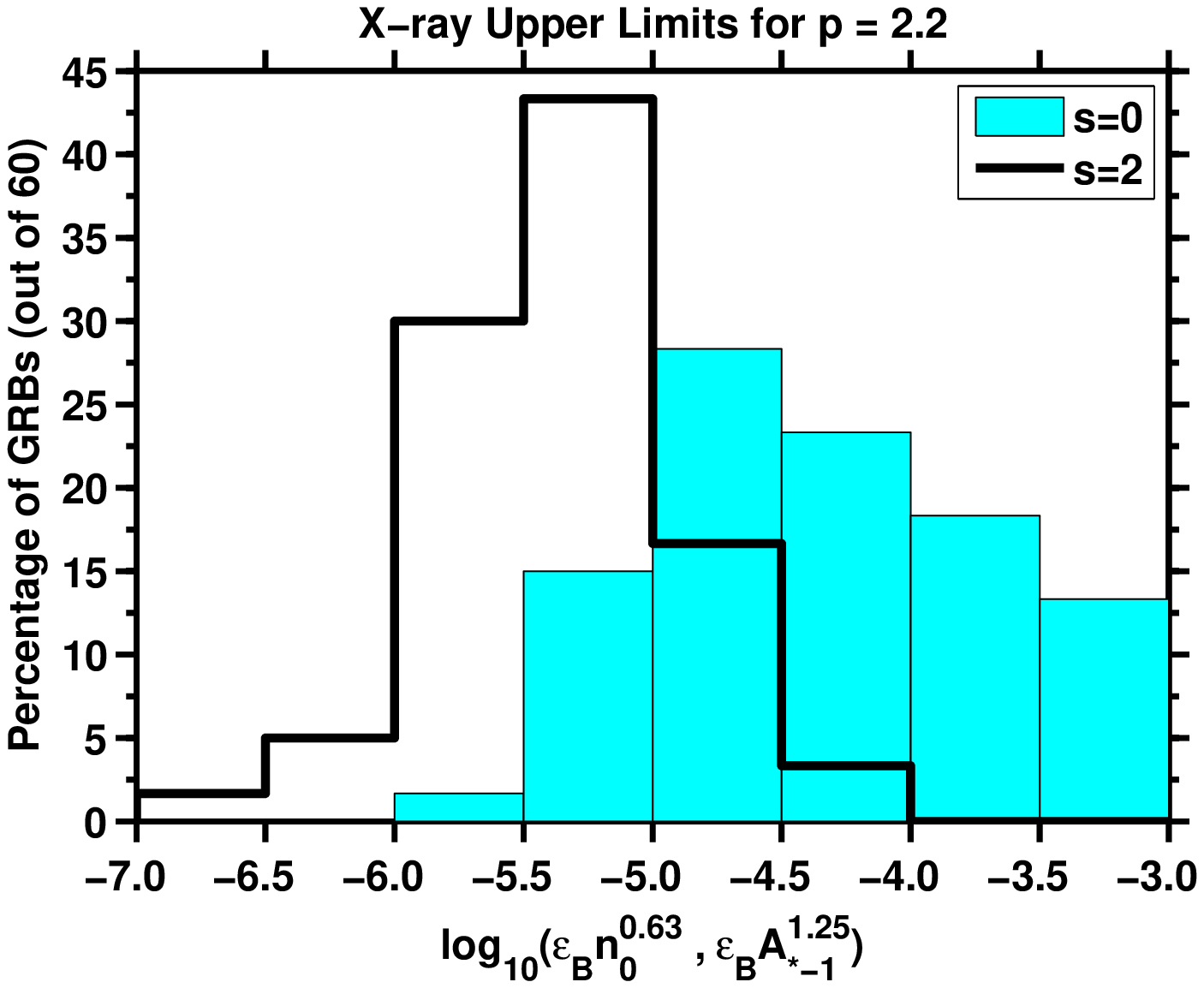} &
\hspace{0mm}
\includegraphics[scale=0.6]{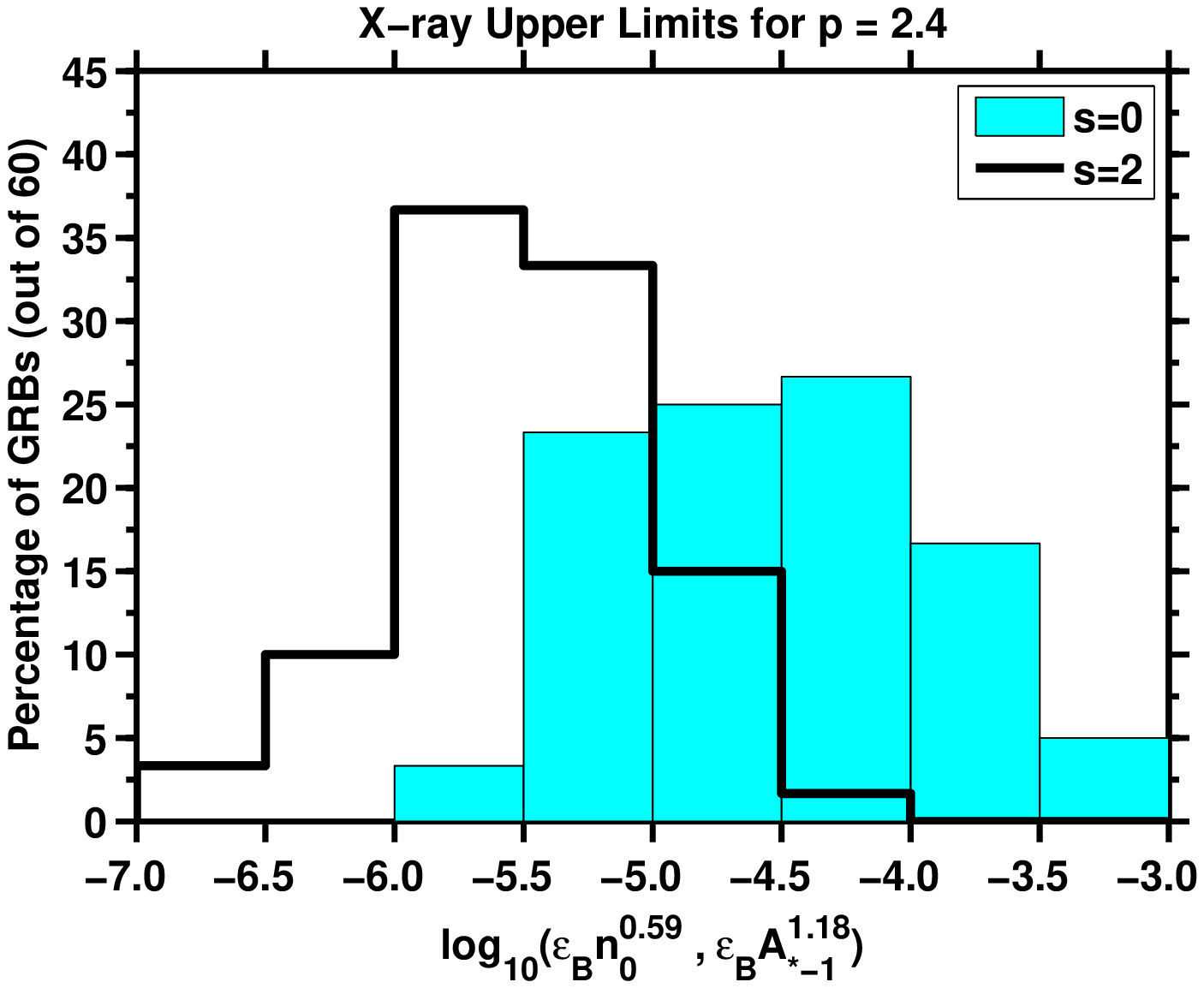}
\end{array}
\\
\hspace{-10mm}
\includegraphics[scale=0.6]{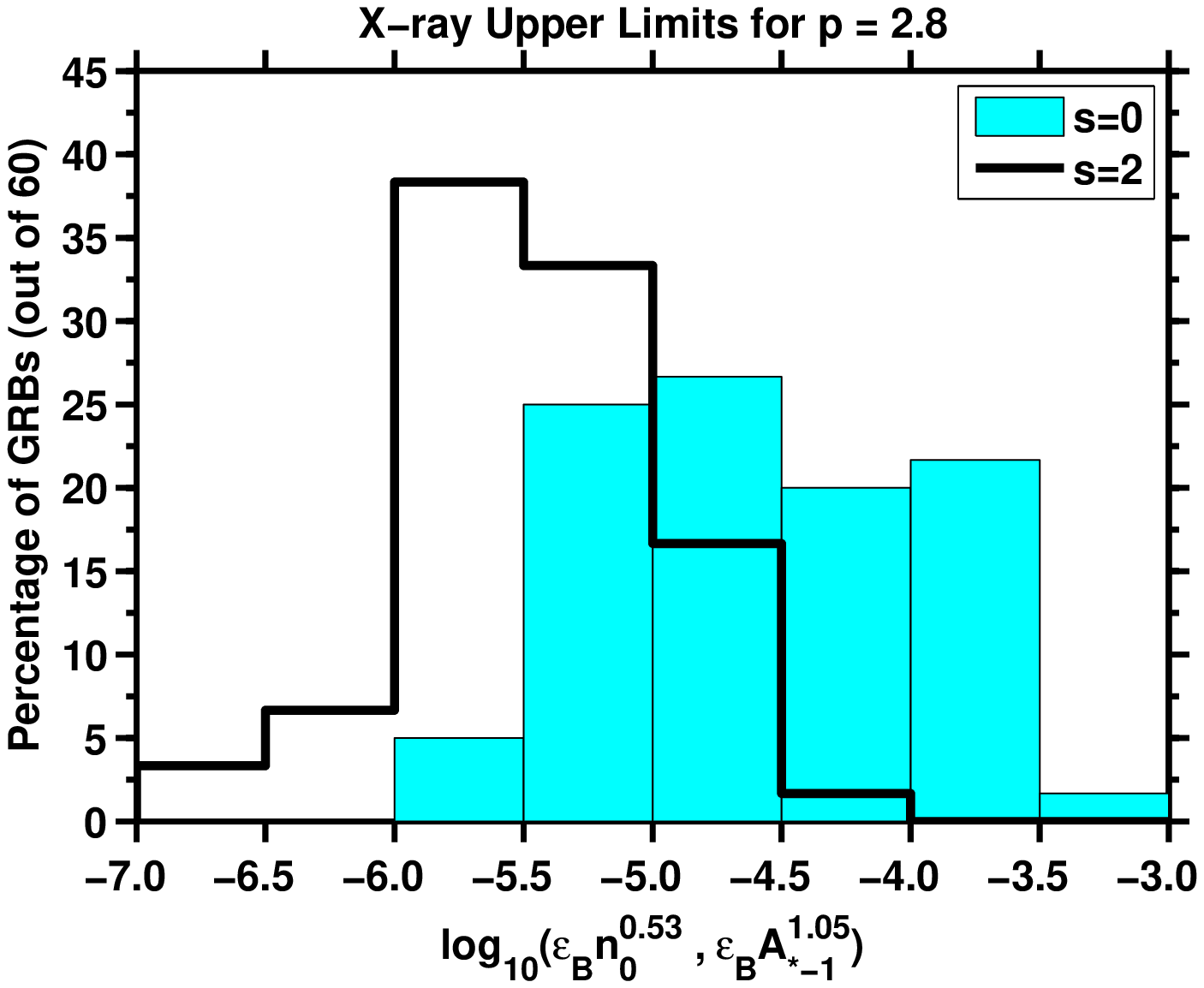}
\end{array}$
\end{center}
\caption{The \textit{Top-Left}, \textit{Top-Right}, and
\textit{Bottom} panels show the histograms of 
upper limits assuming $p=2.2$, $p=2.4$, 
and $p=2.8$ respectively, for all of the GRBs in our X-ray sample (obtained with Equation
\ref{x-ray_constraint}). The filled-in (un-filled) histograms show
upper limits on the quantity $\epsilon_{B} n_{0} ^{2/(p+1)}$ ($\epsilon_{B} A_{*-1} ^{4/(p+1)}$) for
$s=0$ ($s=2$) assuming all the GRBs in our X-ray sample are described
by a constant density (wind) medium. \label{x-ray_epsilon_upper_limit_histograms}}
\end{figure*}

We display the results for the upper limits (from
Equation \ref{x-ray_constraint}) on the quantity 
$\epsilon_{B} n_{0} ^{2/(p+1)}$ ($\epsilon_{B} A_{*-1} ^{4/(p+1)}$) for
$s=0$ ($s=2$)
assuming all GRBs in our X-ray sample have 
$p=2.2$, 2.4, and 2.8 in the Top-Left, Top-Right, and Bottom
panels of Figure \ref{x-ray_epsilon_upper_limit_histograms},
respectively. Two histograms are shown in each 
panel, one for $s=0$ and the other for $s=2$. 
Table \ref{epsilon_B_results_table} shows the mean and median
 upper limits on the quantity 
$\epsilon_{B} n_{0} ^{2/(p+1)}$ ($\epsilon_{B} A_{*-1} ^{4/(p+1)}$) for
$s=0$ ($s=2$) for each histogram. 
\begin{table}[t]
\centering 
\begin{small}
\begin{tabular}{| l | l l l l| }
\hline
X-ray & & & &	\\
$(s=0)$ & $p=2.2$ & $p=2.4$ & $p=2.8$ &  \\
\hline
Mean	&	$ 1.1 \times 10^{-4} $	&	$ 7.2 \times 10^{-5} $ & $ 6.1 \times 10^{-5} $	& \\
Median	&	$ 4.2 \times 10^{-5} $	&	$ 2.8 \times 10^{-5} $ &	$ 2.0 \times 10^{-5} $	& \\	
\hline
X-ray & & & &	\\
$(s=2)$ & $p=2.2$ & $p=2.4$ & $p=2.8$ &  \\
\hline
Mean	&	$ 7.9 \times 10^{-6} $	&	$ 5.5 \times 10^{-6} $ &	$ 5.6 \times 10^{-6} $	& \\
Median	&	$ 4.0 \times 10^{-6} $	&	$ 3.2 \times 10^{-6} $ &	$ 3.4 \times 10^{-6} $	& \\
\hline
\hline
Opt.	&		&		&	&	\\
$(s=0)$ & $p=2.2$ & $p=2.4$ & $p=2.8$ & $p$ from $\alpha_{\mathrm{O}}$	\\
\hline
Mean	&	$ 3.5 \times 10^{-4}$	&	$ 1.0 \times 10^{-4} $	&	$ 2.9 \times 10^{-5} $ & $ 9.5 \times 10^{-5} $	\\
Median	&	$ 1.1 \times 10^{-4}$	&	$ 3.3 \times 10^{-5} $ &	$ 5.5 \times 10^{-6} $ & 	$ 2.4 \times 10^{-5} $	\\
\hline
Opt.	&		&		&	&	\\
$(s=2)$ & $p=2.2$ & $p=2.4$ & $p=2.8$ & 	\\
\hline
Mean	&	$ 7.2 \times 10^{-5} $	&	$ 2.4 \times 10^{-5} $	&	$ 7.5 \times 10^{-6} $ &	\\
Median	&	$ 1.2 \times 10^{-5} $	&	$ 3.9 \times 10^{-6} $	&	$ 8.8 \times 10^{-7} $ &	\\
\hline																									
\end{tabular}
\end{small}
\caption{Mean and median $\epsilon_{B}$ values for the
X-ray (upper limits on $\epsilon_{B}$) and optical (measurements of
$\epsilon_{B}$) histograms shown in Figures
\ref{x-ray_epsilon_upper_limit_histograms} and 
\ref{optical_epsilon_B_measurement_histograms}. 
The section labeled ``X-ray $(s=0)$'' (``X-ray $(s=2)$'') shows the mean and
median $\epsilon_{B}$ upper limits assuming a
constant density (wind) medium with a standard $n=1 \mbox{ cm}^{-3}$
($A_{*} = 0.1$). The columns show the
value of $p$ that was assumed.
The section labeled ``Opt. $(s=0)$''
(``Opt. $(s=2)$'') shows the mean and median $\epsilon_{B}$ measurements assuming a
constant density (wind) medium with a standard $n = 1 \mbox{ cm}^{-3}$
($A_{*} = 0.1$). The columns show the
value of $p$ that was assumed. The column labeled ``$p$ from
$\alpha_{O}$'' shows the mean and median $\epsilon_{B}$
measurements with $p$ determined from $\alpha_{O}$. There are 60 GRBs
in our X-ray sample and 35 GRBs in our optical sample. 
\label{epsilon_B_results_table}}
\end{table}
For the remainder of this section,
we assume a standard $n_{0} = 1$ ($A_{*-1} = 1$) for $s=0$ ($s=2$)
when discussing our results for the $\epsilon_{B}$ upper limits. 

The $\epsilon_{B}$ upper
limit histograms show a wide distribution. For a constant density
(wind) medium, all the histograms show a distribution ranging
from $\sim 10^{-6} - 10^{-3}$ ($\sim 10^{-7} - 10^{-4}$). 
For a constant density (wind) medium, the mean and median 
$\epsilon_{B}$ upper limit values are  $\sim \mbox{ few} \times
10^{-5}$ ($\sim \mbox{ few} \times 10^{-6}$). 
Assuming a different value of $p$ does not 
have a significant effect on the distributions of the $\epsilon_{B}$ upper limits for our X-ray sample. For
both the $s=0$ and $s=2$ cases, when changing $p$, the mean and median
$\epsilon_{B}$ upper limit values change by
less than a factor $\sim 2$. 
Although previous afterglow studies 
also showed a wide distribution for $\epsilon_{B}$ (Figure
\ref{literature_compilation_histograms}), our distribution of $\epsilon_{B}$ upper
limits is shifted towards lower values. Unlike 
Figure \ref{literature_compilation_histograms}, which shows that many GRBs have been
reported to have $\epsilon_{B} \sim 10^{-3} - 10^{-1}$, none of our
histograms of $\epsilon_{B}$ upper limits show an $\epsilon_{B}$ upper
limit larger than $10^{-3}$.

We now discuss how our assumptions on the afterglow parameters can
affect the distribution of $\epsilon_{B}$ upper limits. For this
discussion, we will take a typical $p=2.4$; 
Equation \ref{epsilon_B_afterglow_parameters}
shows how the $\epsilon_{B}$ upper limit depends on the
other afterglow parameters. The strongest dependence is on
$\epsilon_{e}$, which is raised to the power of $-1.65$. However, as
we displayed in Figure \ref{literature_compilation_histograms}, according to
previous studies, the distribution of
$\epsilon_{e}$ values is narrow, with the 
$\epsilon_{e}$ values ranging only over one order of magnitude. 
In addition, $\sim 62\%$ of the bursts have 
$\epsilon_{e} \sim 0.1 - 0.3$. From Figure \ref{literature_compilation_histograms},
a likely error in $\epsilon_{e}$ from our assumed $\epsilon_{e} = 0.2$
is a factor $\sim 2$. From Equation \ref{epsilon_B_afterglow_parameters},
an error in $\epsilon_{e}$ by a factor $\sim 2$ will only lead to an error
in the $\epsilon_{B}$ upper limit by a factor $\sim 3$. 
For a constant density (wind) medium, the $\epsilon_{B}$ upper
limit depends on $E$ as $E^{-1.59}$ ($E^{-1}$). We assumed an
efficiency of $\sim 20\%$ in the conversion of kinetic energy to prompt
gamma-ray radiation. Recent studies have found higher values for the
efficiency (\citealt{granot_et_al_2006}, \citealt{zhang_et_al_2007},
see however, \citealt{fan_and_piran_2006}). In \citealt{zhang_et_al_2007}, the mean 
(median) efficiency they reported 
is $\sim 37\%$ ($\sim 32\%$). Taking the
efficiency to be $\sim 30\%-40\%$ instead of $\sim 20\%$ would lead to
an error in $E$ by a factor $\sim 2-3$. From Equation 
\ref{epsilon_B_afterglow_parameters}, an error in $E$ by a factor
$\sim 2-3$ would lead to an error in the $\epsilon_{B}$ upper limit by
a factor $\sim 3-6$ ($\sim 2-3$) for a constant density (wind) medium.
Lastly, the largest source of uncertainty for the
$\epsilon_{B}$ upper limits is the density, since it 
has been reported to have a range 
$\sim 10^{-3} \mbox{ cm}^{-3} - 10^{2} \mbox{ cm}^{-3}$.
For a constant density (wind) medium, the $\epsilon_{B}$ upper
limit depends on the density as $n^{-0.59}$ ($A_{*} ^{-1.18}$). For
$s=0$ ($s=2$), we assumed a standard $n = 1 \mbox{ cm}^{-3}$ 
($A_{*} = 0.1$). An error in the density by a factor $\sim 10^{3}$
($\sim 10^{2}$) will lead to an error in the $\epsilon_{B}$ upper
limit by a factor $\sim 60$ ($\sim 230$).

In summary, the expected errors in $\epsilon_{e}$ and $E$ of a factor
$\sim 2-3$ will not change the $\epsilon_{B}$ upper limits by an order
of magnitude. On the other hand, the density is a very uncertain
parameter and an error in the density by $\sim 2-3$ orders of
magnitude will lead to an error in the $\epsilon_{B}$ upper limits by 
$\sim 1-2$ orders of magnitude. 

Two additional parameters that can affect our
  $\epsilon_{B}$ upper limits are: 1. $\xi$, the fraction of electrons
  accelerated to a power-law distribution and 2. $f$, which is a factor
  that takes into account the degeneracy for a set of afterglow parameters.
  For a set of parameters $E$, $n$, $\epsilon_{e}$, $\epsilon_{B}$, $\xi$ producing the
  observed external-forward shock flux, another set of primed
  parameters $E' = E/f$, $n' = n/f$, $\epsilon_{e}' = f \epsilon_{e}$,
 $\epsilon_{B}' = f \epsilon_{B}$, $\xi' = f \xi$ can also produce the
 observed external-forward shock flux
 \citep{eichler_and_waxman_2005}. Afterglow studies usually assume
 $\xi = f = 1$ for simplicity but $\xi \le 1$ and $m_{e}/m_{p} \le f \le
 1$ \citep{eichler_and_waxman_2005}, where $m_{e}$ ($m_{p}$) is the
 electron (proton) mass. The external-forward shock flux
 depends on $\xi$ and $\epsilon_{B}$ as $f_{\nu} \propto \xi^{2-p}
 \epsilon_{B} ^{(p+1)/4}$ \citep{leventis_et_al_2012}. From this
 dependence, we find that the $\epsilon_{B}$ upper limit
 $\propto \xi^{4(p-2)/(p+1)}$. Thus, including $\xi$ will decrease 
 the values of our $\epsilon_{B}$ upper
 limits. Values of $\xi$ have not been determined for GRB
 external-forward shocks so we cannot quantify by how much 
 the $\epsilon_{B}$ upper limit values will decrease. Including $f$
 will also decrease the values of the $\epsilon_{B}$ upper limits
 since $\epsilon_{B}' = f \epsilon_{B}$ and $f < 1$. Like $\xi$,
 values of $f$ have also not been determined from afterglow observations.
 The largest effect $f$ can have on the $\epsilon_{B}$ upper limit
 values is decrease them by a factor of $m_{p}/m_{e} \sim 2000$. For
 the remainer of this paper, we will be conservative and continue to
 assume $\xi = f = 1$, but we should keep in mind that considering 
$\xi$ and $f$ will decrease the values of the $\epsilon_{B}$ upper limits.


\section{Measurement Of $\epsilon_{B}$ With Optical Light Curves}\label{optical_section}

\subsection{$\epsilon_{B}$ Determination With Optical Data}\label{optical_epsilon_B_arguments}

The light curves we consider for our optical sample
decline with a temporal decay index $\alpha \sim 1$ from early times,
$\sim 10^{2} - 10^{3} \mbox{ sec}$, as expected for the
external-forward shock emission (see Section \ref{optical_afterglow_parameter_assumption}). 
Since the light curves of these bursts are likely dominated by the external-forward shock, this means
that the observed optical flux is an actual measurement of
the external-forward shock flux, that is\footnote{At late times, $\sim 10^3 - 10^5$ sec, many X-ray light curves decline with $\alpha \sim 1$: 
the ``normal'' decline.  If this segment arises from the external-forward shock, then $\epsilon_B$ 
can be determined as described in this Section for our optical sample.  However, this is not 
straightforward since energy injection (as evidenced by the plateau
phase) should be considered. Also, one can attempt to use the upper limit on $\epsilon_B$, found
in Section \ref{section_3}, to calculate a lower limit on 
$\nu_c$ during the normal decline and compare this to the observed
spectral regime ($\mbox{1 keV} < \nu_{c}$ or $\mbox{1 keV} >
\nu_{c}$). However, there are two difficulties with this $\nu_{c}$ consistency
check: 1. several ``closure relations'' are simultaneously
satisfied within 2-$\sigma$ for most of our sample
(due to large errors in the spectral and temporal indices at late
times, which can be found in \citet[][]{evans_et_al_2007,evans_et_al_2009}).  
2. for the bursts that uniquely satisfy one closure relation, $\nu_c$
cannot be determined precisely since
Klein-Nishina suppression is weaker at late times and $\nu_{c}$ depends
strongly on $n$ when Compton-$Y$ is considered.}
\begin{equation}
f_{\mathrm{obs}} = f_{\mathrm{ES}} (E , n , \epsilon_{e} , \epsilon_{B} , p, s) . \label{symbolic_optical_expression}
\end{equation}
Later in this section we will use this condition to determine 
$\epsilon_{B}$ for the bursts in our optical
sample. We want to stress that we determine $\epsilon_{B}$ for the
optical sample. This is in contrast to the X-ray sample, which
only allowed us to determine an upper limit on $\epsilon_{B}$.
\subsection{The Optical Sample}\label{optical_afterglow_parameter_assumption}
Our optical sample consists of 35 GRBs. $33/35$ of the bursts 
triggered \textit{Swift} and the remaining two bursts,
050502A and 080603A, were detected by INTEGRAL (INTErnational Gamma-Ray
Astrophysics Laboratory, \citealt{winkler_et_al_2003}). Table \ref{optical_properties_table} shows properties 
of our optical sample. With a few exceptions, most of the
GRBs in our optical sample have a known redshift\footnote{The 
redshifts were taken from NASA's \textit{Swift} GRB Table and Lookup
website (exceptions are GRB 071003 \citep{perley_et_al_2008} and GRBs
050502A and 071025 (J. Greiner's website)). Three 
GRBs (050721, 070420, 060111B) in our optical sample do not have a known redshift. For GRB 050721 
and GRB 070420, we assumed the average redshift of 2.5 for \textit{Swift} GRBs 
\citep{gehrels_et_al_2009}. The redshift for GRB 060111B was approximated as 2 in 
\citet{stratta_et_al_2009}.}. 
%
%
%
\begin{small}
\begin{longtable*}{lcccccccccc}
\caption{Optical Sample Properties \label{optical_properties_table}}\\
\hline
GRB & 
$z$  &
$d_{L28}$  &
Fluence  & 
$E ^{\gamma} _{\mathrm{iso, }52}$  & 
$\alpha_{O}$  & 
Ref.  & 
$t_{2}$  & 
$f_{2 \mathrm{eV}}$ &
$\mbox{log}_{10}(\epsilon_{B})$ \\
 & 
 &
 &
$[\times 10^{-6} \mbox{ ergs}/\mbox{cm}^{2}]$ & 
 & 
 & 
 & 
 & 
[mJy] &
($s=0$) \\
\hline
\endfirsthead
\caption{Optical Sample Properties (Continued)}\\
\hline
GRB & 
$z$  &
$d_{L28}$  &
Fluence  & 
$E ^{\gamma} _{\mathrm{iso, }52}$  & 
$\alpha_{O}$  & 
Ref.  & 
$t_{2}$  & 
$f_{2 \mathrm{eV}}$ &
$\mbox{log}_{10}(\epsilon_{B})$ \\
 & 
 &
 &
$[\times 10^{-6} \mbox{ ergs}/\mbox{cm}^{2}]$ & 
 & 
 & 
 & 
 & 
[mJy] &
($s=0$) \\
\hline
\endhead
\hline 
\endfoot
\hline
\endlastfoot
\textbf{050401}	&	2.9	&	7.67	&	8.22	&	15.55	&	$	0.80 \pm 0.03	$	&	[1]	&	0.72	&	0.3	&	-5.5	\\
\textit{050502A}	&	3.793	&	10.59	&	1.4	&	4.12	&	$	1.16 \pm 0.03	$	&	[2]	&	1	&	5	&	-4.5	\\
050525A	&	0.606	&	1.10	&	15.3	&	1.45	&	$	1.12 \pm 0.05	$	&	[3]	&	34.56	&	0.5	&	-4.3	\\
\textbf{050721}	&	2.5	&	6.40	&	3.62	&	5.32	&	$	1.29 \pm 0.06	$	&	[4]	&	20	&	0.2	&	-5.0	\\
050730	&	3.97	&	11.19	&	2.38	&	7.53	&	$	0.89 \pm 0.05	$	&	[17]	&	7.5	&	0.57	&	-3.9	\\
050802	&	1.71	&	4.00	&	2.00	&	1.49	&	$	0.82 \pm 0.03	$	&	[1]	&	3.6	&	0.5	&	-3.5	\\
051111	&	1.55	&	3.54	&	4.08	&	2.52	&	$	1.00 \pm 0.02	$	&	[2]	&	30	&	0.4	&	-3.8	\\
\textbf{051221A}	&	0.5465	&	0.97	&	1.15	&	0.09	&	$	0.96 \pm 0.03	$	&	[5]	&	100	&	0.02	&	-3.4	\\
\textbf{060111B}	&	2	&	4.86	&	1.60	&	1.58	&	$	1.18 \pm 0.05	$	&	[6]	&	2	&	0.4	&	-5.2	\\
\textbf{060210}	&	3.91	&	10.99	&	7.66	&	23.65	&	$	1.03 \pm 0.06	$	&	[2]	&	10	&	0.1	&	-6.0	\\
\textbf{060418}	&	1.49	&	3.37	&	8.33	&	4.78	&	$	1.13 \pm 0.02	$	&	[7]	&	2	&	8	&	-4.6	\\
\textbf{060607A}	&	3.082	&	8.25	&	2.55	&	5.34	&	$	1.20 \pm 0.03	$	&	[7]	&	2	&	10	&	-4.2	\\
060904B	&	0.703	&	1.32	&	1.62	&	0.21	&	$	1.00 \pm 0.18	$	&	[17]	&	5.5	&	0.58	&	-3.5	\\
060908	&	2.43	&	6.18	&	2.80	&	3.91	&	$	1.05^{+0.03} _{-0.03}	$	&	[8]	&	2	&	2	&	-4.5	\\
\textbf{060927}	&	5.6	&	16.81	&	1.13	&	6.08	&	$	1.21 \pm 0.06	$	&	[2]	&	0.5	&	2	&	-5.5	\\
061007	&	1.26	&	2.74	&	44.4	&	18.48	&	$	1.70 \pm 0.02	$	&	[7]	&	2	&	50	&	-6.0	\\
061110B	&	3.44	&	9.42	&	1.33	&	3.34	&	$	1.64 \pm 0.08	$	&	[2]	&	20	&	0.02	&	-5.9	\\
\textbf{061121}	&	1.314	&	2.88	&	13.7	&	6.19	&	$	0.82 \pm 0.02	$	&	[7]	&	4	&	0.5	&	-4.7	\\
061126	&	1.159	&	2.47	&	6.77	&	2.39	&	$	0.89 \pm 0.02	$	&	[2]	&	10	&	0.2	&	-4.6	\\
070318	&	0.84	&	1.65	&	2.48	&	0.46	&	$	0.96 \pm 0.03	$	&	[7]	&	20	&	0.2	&	-3.7	\\
070411	&	2.954	&	7.84	&	2.70	&	5.27	&	$	0.92 \pm 0.04	$	&	[2]	&	20	&	0.07	&	-4.7	\\
070420	&	2.5	&	6.40	&	14.0	&	20.56	&	$	0.81 \pm 0.04	$	&	[2]	&	3	&	0.8	&	-4.7	\\
\textbf{070714B}	&	0.92	&	1.85	&	0.72	&	0.16	&	$	0.83 \pm 0.04	$	&	[2]	&	10	&	0.03	&	-3.7	\\
071003	&	1.6	&	3.69	&	8.3	&	5.45	&	$	1.466 \pm 0.006	$	&	[9]	&	0.6	&	20	&	-5.7	\\
071025	&	5.2	&	15.41	&	6.5	&	31.26	&	$	1.27 \pm 0.04	$	&	[10]	&	20	&	0.02	&	-6.8	\\
071031	&	2.692	&	7.00	&	0.9	&	1.5	&	$	0.97 \pm 0.06	$	&	[11]	&	10.5	&	0.4	&	-3.4	\\
071112C	&	0.823	&	1.61	&	3.00	&	0.53	&	$	0.95 \pm 0.02	$	&	[12]	&	10.5	&	0.003	&	-6.3	\\
\textit{080603A}	&	1.688	&	3.94	&	1.1	&	0.80	&	$	0.99 \pm 0.07	$	&	[13]	&	30	&	0.1	&	-3.6	\\
\textbf{080607}	&	3.036	&	8.10	&	24.0	&	49.04	&	$	1.65	$	&	[14]	&	3	&	0.2	&	-8.0	\\
\textbf{080721}	&	2.591	&	6.68	&	12.0	&	18.74	&	$	1.22 \pm 0.01	$	&	[5]	&	3	&	10	&	-5.0	\\
080810	&	3.35	&	9.13	&	4.60	&	11.06	&	$	1.23 \pm 0.01	$	&	[7]	&	3	&	30	&	-3.9	\\
080913	&	6.7	&	20.72	&	0.56	&	3.92	&	$	1.03 \pm 0.02	$	&	[15]	&	10	&	0.02	&	-5.2	\\
\textbf{081008}	&	1.967	&	4.76	&	4.30	&	4.13	&	$	0.96 \pm 0.03	$	&	[16]	&	2	&	3	&	-4.1	\\
090313	&	3.375	&	9.21	&	1.40	&	3.41	&	$	1.25 \pm 0.08	$	&	[17]	&	20	&	2	&	-3.4	\\
\textbf{090418A}	&	1.608	&	3.71	&	4.60	&	3.05	&	$	1.21 \pm 0.04	$	&	[7]	&	2	&	0.8	&	-5.5	\\
\hline
\caption{\footnotesize{The bursts in bold are also part of our X-ray
sample and the two bursts in italics were detected by INTEGRAL, instead
of \textit{Swift}. The redshift and the corresponding luminosity
distance in units of $10^{28}$ cm, $d_{L28}$, are shown
in the second and third columns, respectively. The fluence, in units of $10^{-6} \mbox{ergs}/\mbox{cm}^{2}$, is shown in the fourth
column. In the next column we show the isotropic equivalent energy
released in gamma-rays during the prompt emission in the units of $10^{52} \mbox{ ergs}$,
$E ^{\gamma} _{\mathrm{iso, }52}$. The temporal decay of the optical
light curve, $\alpha_{O}$, is shown in the sixth column. The reference where we found each
optical light curve and $\alpha_{O}$ is shown in the seventh column. 
The time in units of $10^{2}$ seconds, and the
flux in mJy of the data point we used to determine $\epsilon_{B}$ are
shown in the next two columns. For 050730 and 060904B, we display
the time and flux at the peak of the optical light curve
that are given in \citet{melandri_et_al_2010} (we could not find a
optical light curve in units of specific flux in the literature for
these two bursts). In the last column we show the
$\epsilon_{B}$ measurements for $n = 1 \mbox{ cm}^{-3}$ and $p$
determined from $\alpha_{O}$ (See Bottom-Right panel of
Figure \ref{optical_epsilon_B_measurement_histograms}). References for light curves and
$\alpha_{O}$:
$[1] = $ \citet{panaitescu_et_al_2006} 
$[2] = $ \citet{melandri_et_al_2008} 
$[3] = $ \citet{panaitescu_et_al_2007} 
$[4] = $ \citet{antonelli_et_al_2006} 
$[5] = $ \citet{schulze_et_al_2011} 
$[6] = $ \citet{stratta_et_al_2009} 
$[7] = $ \citet{panaitescu_and_vestrand_2011} 
$[8] = $ \citet{covino_et_al_2010} 
$[9] = $ \citet{perley_et_al_2008} 
$[10] = $ \citet{perley_et_al_2009} 
$[11] = $ \citet{kruhler_et_al_2009}
$[12] = $ \citet{uehara_et_al_2010} 
$[13] = $ \citet{guidorzi_et_al_2011}
$[14] = $ \citet{perley_et_al_2011} 
$[15] = $ \citet{greiner_et_al_2009} 
$[16] = $ \citet{yuan_et_al_2010} 
$[17] = $ \citet{melandri_et_al_2010}.
 }}
\end{longtable*}
\end{small}
%
%
%

With the exception
of the only short GRB in our optical sample, GRB 051221A, all the
optical light curves in our sample decline before 3500 seconds. Considering early times has the
advantage of minimizing possible energy injection. Our optical sample can be separated into 4 different
subgroups, depending on the temporal behavior of the light curve before the $\alpha \sim 1$ decay as follows.
\begin{enumerate}
\item \underline{Light Curves With A Peak At Early Times}:
The light curves of this subgroup are characterized by a
power law rise, reaching a peak, and 
then a power law decline with $\alpha_{O} \sim 1$ ($\sim$43\% of our sample). The peak of the light curve 
is believed to be due to the deceleration time. For the bursts in this subgroup, 
we show the temporal decay of the optical light curve after the
peak and the flux and the time of the second data point after the
peak in Table \ref{optical_properties_table}. We take the second data point to
be confident that the optical light curve is declining. 
\item \underline{Single Power Law Decay From Early Times:} In this subgroup,
the optical light curve shows a decline as a single power law with
$\alpha \sim 1$ from the
beginning of the observations ($\sim$ 40\% of our sample). We display the temporal decay of the
optical light curve and the time and the flux of the second
data point observed in Table \ref{optical_properties_table}.
\item \underline{Optical Light Curves With Plateaus At Early Times:}
The optical light curves of 3 bursts in our optical sample (GRBs
050525A, 060210, and 070411) display plateaus
at early times. The plateaus in our optical sample are short, with the longest plateau
lasting under 3500 seconds. After the plateau ends, the light
curves of these 3 bursts show a decay $\alpha_{O} \sim 1$, as expected
for the external forward-shock emission. In Table \ref{optical_properties_table}, for
these 3 bursts, we show the temporal decay
after the plateau and the time and the flux of the second data point after the plateau.
\item \underline{Light Curves With Possible Reverse Shock} 
\underline{Emission At Early Times:} 3 GRBs in our optical sample 
(060111B, 060908, and 061126) show possible emission from 
the reverse shock. The light curves in this subgroup show
an initial steep decline at early times, characteristic
of the reverse shock, and then transition to a
more shallow decay of $\alpha \sim 1$ that is 
more typical for the external-forward shock emission. For these GRBs, 
in Table \ref{optical_properties_table} we show the temporal decay of the light
curve and the time and flux
of the second data point after the possible reverse shock emission.
\end{enumerate}

\subsection{Optical External-Forward Shock Spectral Regime and
  Afterglow Parameter Assumptions}\label{optical_spectrum_section}
When referring to the optical band, we will use 2 eV since most of
the light curves in our optical sample are either plotted at 2eV or
were observed in the R filter. As we did with the X-ray sample, we
will only consider the slow cooling ordering of the synchrotron
characteristic frequencies,
$\nu_{a} < \nu_{i} < \nu_{c}$. Because the optical
light curve is declining at the time we are considering, the
optical band must be above $\nu_{i}$ at this time. In Section 
\ref{spectrum_at_the_end_of_the_steep_decline}, we argued that the X-ray band 
is between $\nu_{i}$ and $\nu_{c}$ at the end of the steep decline at
a few 100 sec; therefore, the optical
band must also be in this spectral regime at the early times 
$(\sim 10^{2} - 10^{3})$ sec we are considering. The expression we will
use to determine the optical external-forward shock flux
is also Equation \ref{x-ray_constraint}; however, we will have an equality (instead of an inequality), 
we replace $f_{\mathrm{1 keV, EoSD}}$ with $f_{\mathrm{2 eV}}$
(which represents the specific flux observed at 2 eV), and use
$\nu_{14}$ corresponding to 2 eV. 

The other afterglow parameters are determined as in
Section \ref{X-ray_afterglow_parameter_assumptions}: $\epsilon_{e} =
0.2$ and with $z$ and the fluence\footnote{For the two
bursts detected by INTEGRAL, the fluence in Table \ref{optical_properties_table} is in the
20-200 keV band of the instrument IBIS (Imager on-Board the INTEGRAL
Satellite, \citet{ubertini_et_al_2003}): GRB 050502A
\citep{gotz_and_mereghetti_2005} and GRB 080603A \citep{guidorzi_et_al_2011}.}, we obtain $E^{\gamma} _{\mathrm{iso}}$
and use $E = 5 E^{\gamma} _{\mathrm{iso}}$.
As with our X-ray sample, we will display our $\epsilon_{B}$ results
with $p=2.2$, 2.4, and 2.8. We can also determine $p$ by using the temporal
decay of the optical light curve, $\alpha_{O}$, which is shown in
Table \ref{optical_properties_table} for each burst (optical spectrum is not always
available, so we cannot use the closure relations for the optical
sample). In order to have $p>2$ for all of the bursts in our optical
sample, we only consider a constant density medium when determining
$p$ with $\alpha_{O}$ ($\alpha_{O} = 3(p-1)/4$ for $s=0$ and 
$\alpha_{O} = (3 p -1)/4$ for $s=2$). Lastly, as we did for our X-ray
sample, to keep the density-dependence, we will
plot the quantity $\epsilon_{B} n_{0} ^{2/(p+1)}$ 
($\epsilon_{B} A_{*-1} ^{4/(p+1)}$) for $s=0$ ($s=2$).

\subsection{$\epsilon_{B}$ Results For Optical Sample}\label{optical_epsilon_{B}_results}

\begin{figure*}
\begin{center}$
\begin{array}{c}
\begin{array}{cc}
\hspace{-10mm}
\includegraphics[scale=0.6]{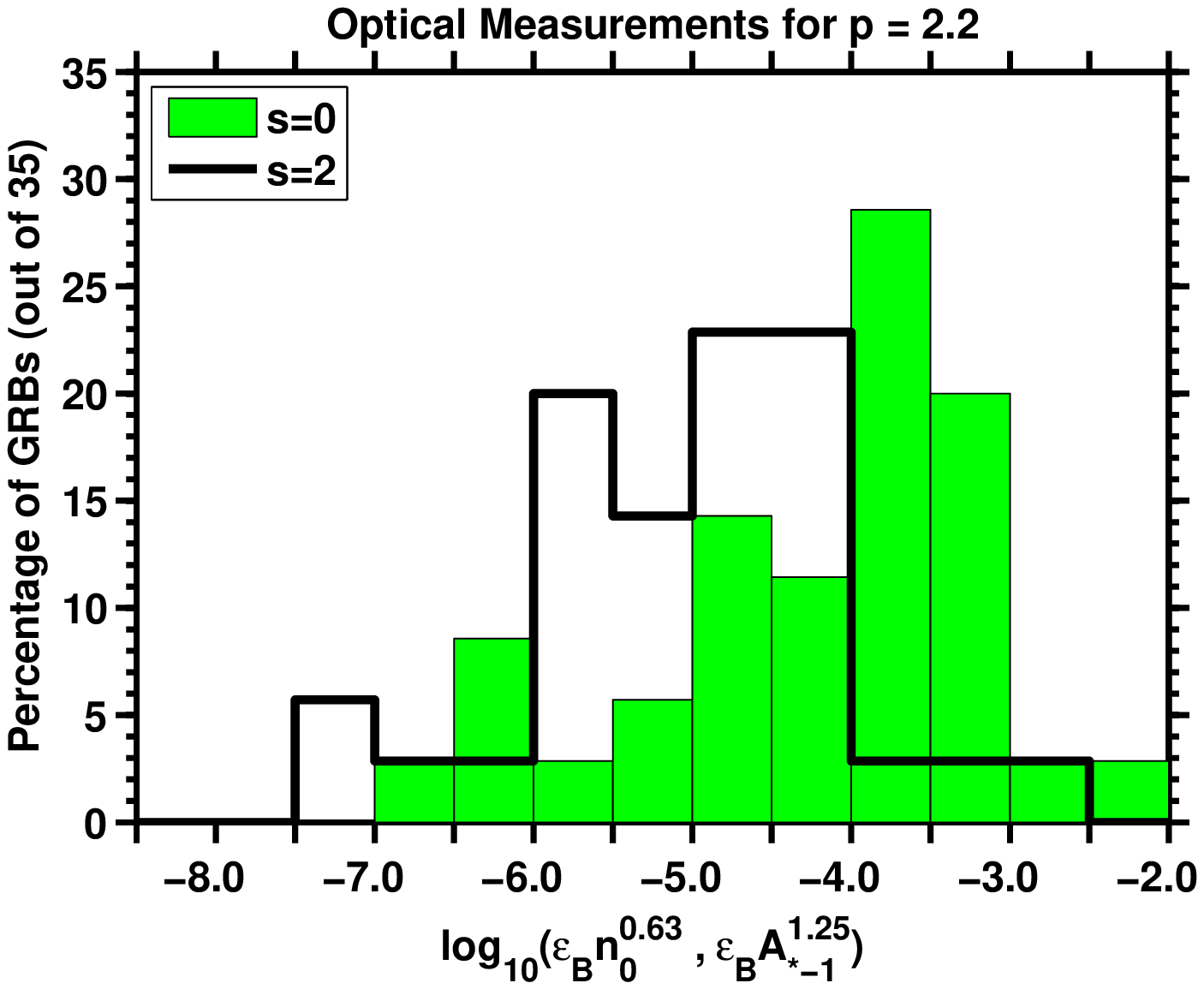} &
\hspace{0mm}
\includegraphics[scale=0.6]{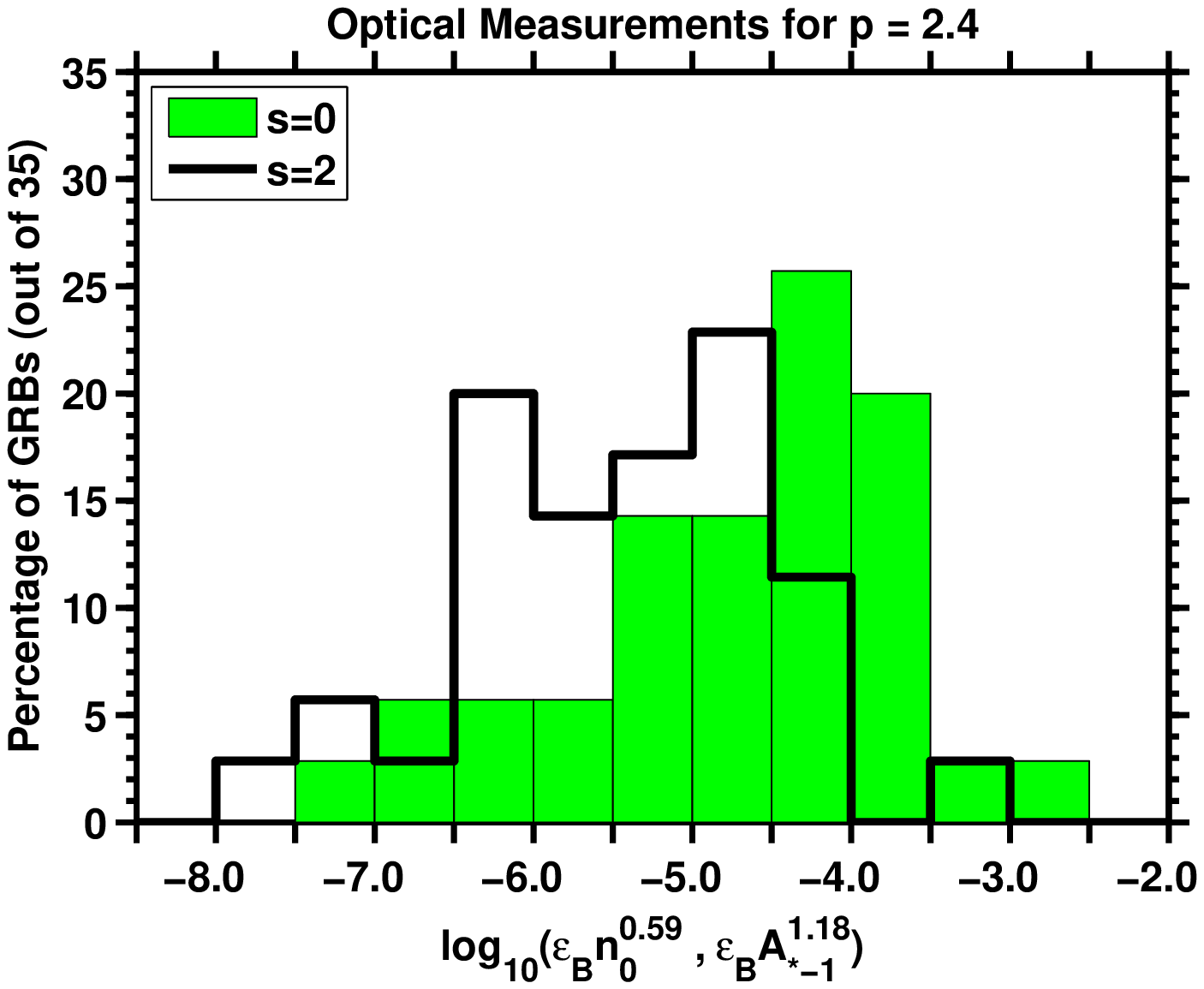}
\end{array}
\\
\begin{array}{cc}
\hspace{-10mm}
\includegraphics[scale=0.6]{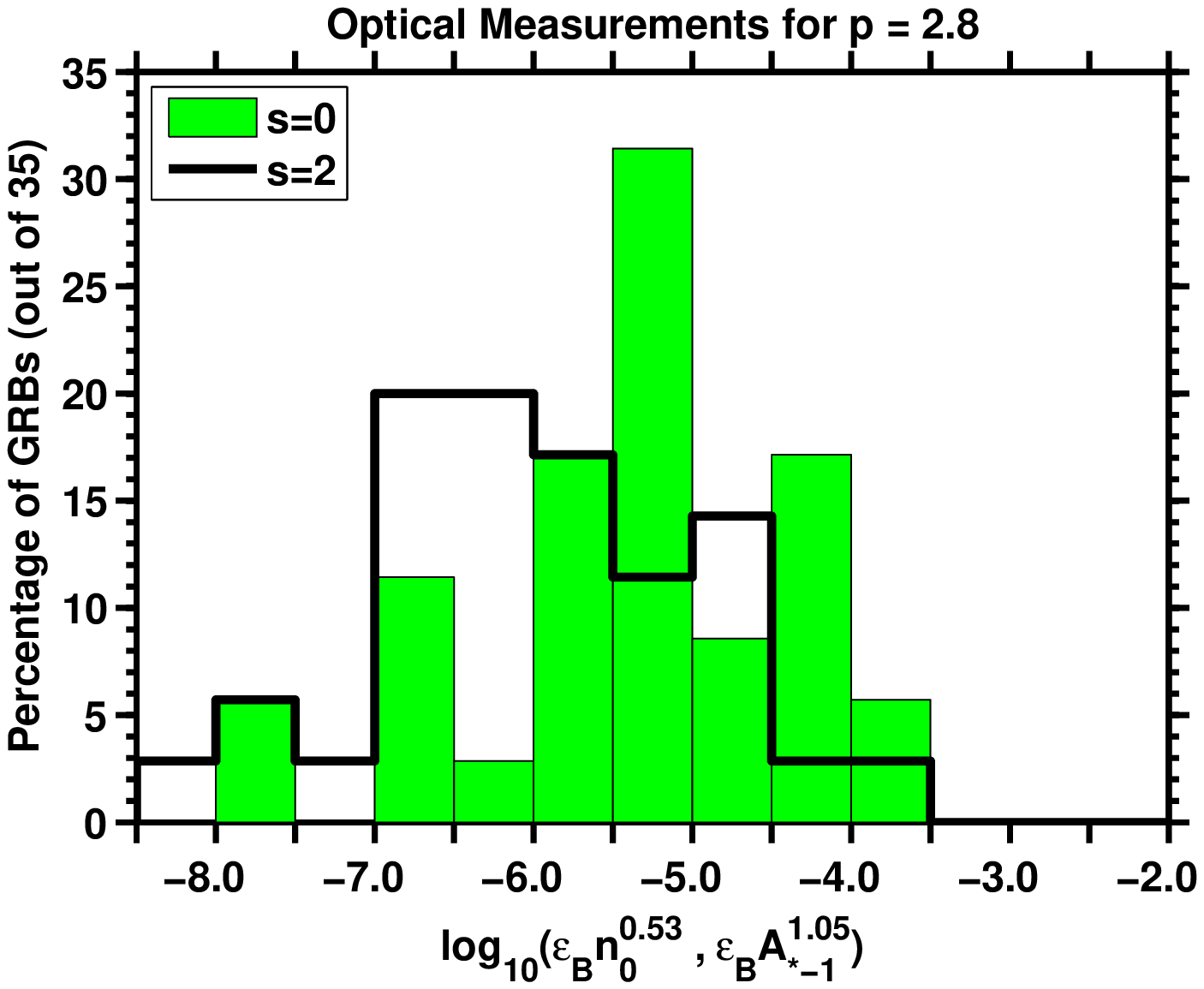} &
\hspace{0mm}
\includegraphics[scale=0.6]{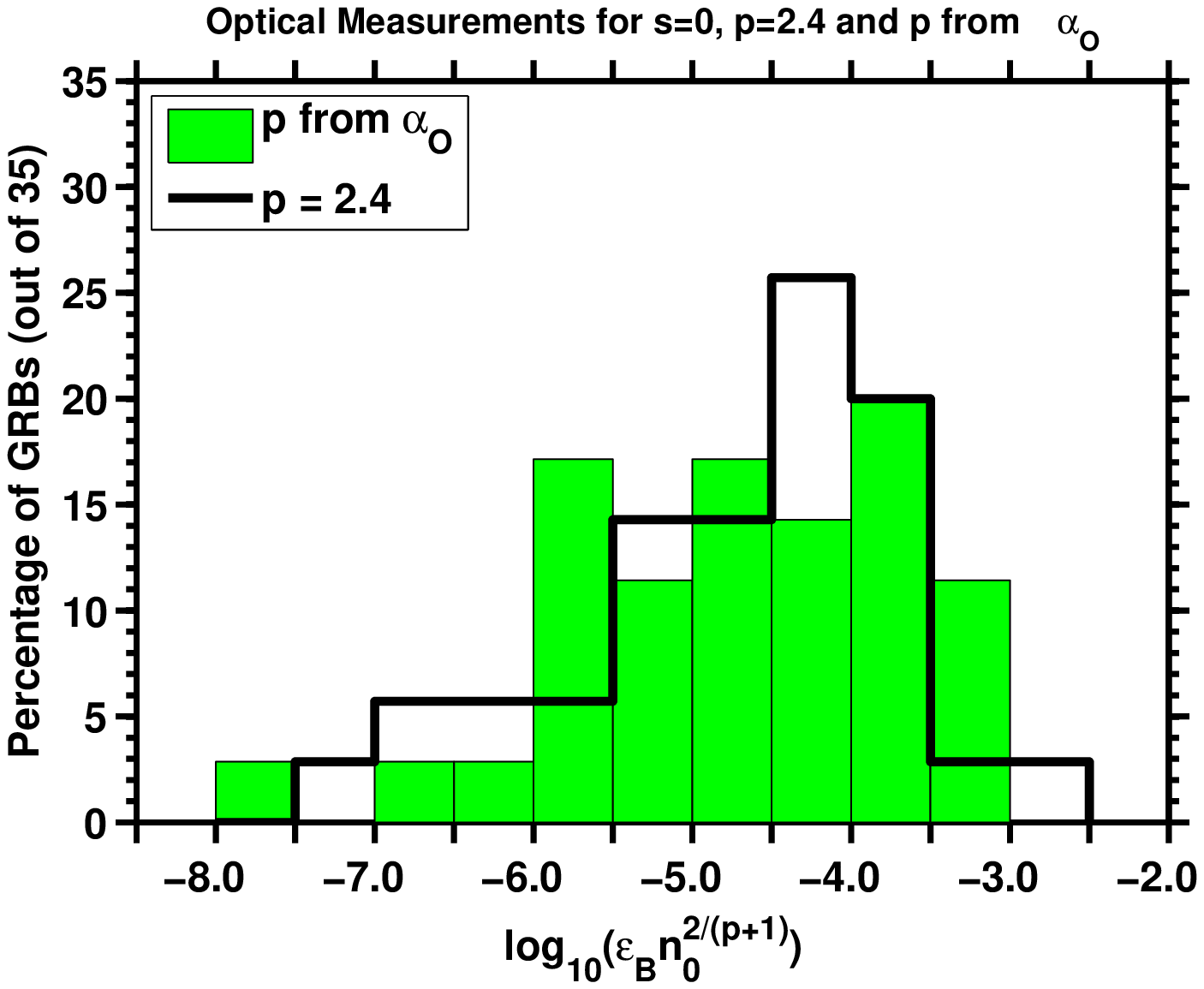}
\end{array}
\end{array}$
\end{center}
\caption{The \textit{Top-Left}, \textit{Top-Right}, and
  \textit{Bottom-Left} panels show the histograms of 
  measurements assuming $p=2.2$, $p=2.4$, 
  and $p=2.8$ respectively for all of the GRBs in our optical
  sample. The filled-in (un-filled) histograms show measurements of
  the quantity $\epsilon_{B} n_{0} ^{2/(p+1)}$ ($\epsilon_{B} A_{*-1}
  ^{4/(p+1)}$) assuming all the GRBs in our optical sample are
  described by a constant density (wind) medium. 
\textit{Bottom-Right Panel:} The filled-in histogram shows the
measurements on the quantity $\epsilon_{B} n_{0} ^{2/(p+1)}$
  with $p$ determined from $\alpha_{O}$. The un-filled histogram shows measurements
  on the quantity $\epsilon_{B} n_{0} ^{2/(p+1)}$, assuming $p = 2.4$ for all of the
  bursts in our optical sample (this histogram was also shown in the
  Top-Right panel). \label{optical_epsilon_B_measurement_histograms}}
\end{figure*}

We display the results for the measurements (from 
Equation \ref{x-ray_constraint}) on the quantity $\epsilon_{B} n_{0}
^{2/(p+1)}$ ($\epsilon_{B} A_{*-1} ^{4/(p+1)}$) for $s=0$ ($s=2$)
assuming all the GRBs in our optical sample have $p=2.2$, 2.4, and 2.8
in the Top-Left, Top-Right, and Bottom-Left panels
of Figure \ref{optical_epsilon_B_measurement_histograms},
respectively. Two histograms are
shown in each panel, one for $s=0$ and the other for $s=2$. We also use 
$\alpha_{O}$ to determine $p$ (assuming $s=0$) and compare the results 
to the ones obtained with $p=2.4$ and $s=0$
(Bottom-Right panel of Figure
\ref{optical_epsilon_B_measurement_histograms}). In Table \ref{epsilon_B_results_table},
we display a summary of the mean and
median values of the measurements of the quantity 
$\epsilon_{B} n_{0} ^{2/(p+1)}$ ($\epsilon_{B} A_{*-1} ^{4/(p+1)}$)
for $s=0$ ($s=2$) for each histogram. For the remainder of this
section, we assume a standard $n_{0} =1$ ($A_{*-1} = 1$) for $s=0$
($s=2$) when discussing our results for the $\epsilon_{B}$
measurements for our optical sample. 

For a constant density (wind) medium, the mean and median $\epsilon_{B}$
measurements are $\sim \mbox{ few} \times 10^{-5}$ 
($\sim \mbox{ few} \times 10^{-6}$)
The mean and median $\epsilon_{B}$ 
measurements only change by a factor of a few when assuming a
different value of $p$. To determine if assuming a standard
$p=2.4$, as opposed to determining $p$ for each burst from $\alpha_{O}$,
significantly affects the distribution of $\epsilon_{B}$
measurements, we compared the two histograms in the Bottom-Right panel of
Figure \ref{optical_epsilon_B_measurement_histograms} with a Kolmogorov-Smirnov (KS) test.
The null hypothesis of the KS test is that the two histograms are
drawn from the same distribution. We test this null hypothesis at the
5\% significance level. 
The KS test confirmed the null hypothesis that the two histograms are
consistent with being drawn from the same distribution.

As with the $\epsilon_{B}$ upper limits
from X-ray data, the mean and median $\epsilon_{B}$ measurements
decrease by about an order of magnitude when assuming a wind medium as
opposed to a constant density medium. Compared to the distribution of $\epsilon_{B}$
upper limits we attained from X-ray data, the $\epsilon_{B}$
measurements from optical data show a much wider distribution. For a
constant density (wind) medium, the $\epsilon_{B}$ measurements
range from $\epsilon_{B} \sim 10^{-8} - 10^{-3}$ 
($\epsilon_{B} \sim 10^{-9} - 10^{-3}$). Also, since we used
the same equation (Equation \ref{x-ray_constraint}) to find both the
upper limits on $\epsilon_{B}$ with X-ray data and the $\epsilon_{B}$
measurements with optical data, the discussion at the end 
of Section \ref{X-ray_results} on how the uncertainty in
the afterglow parameters and the parameters $\xi$ and $f$ can 
affect the distribution of $\epsilon_{B}$ upper limits also applies to the
distributions of $\epsilon_{B}$ measurements we presented in this
section. In addition, since $\xi$ and $f$ are less than unity,
  including these two parameters will mean that our $\epsilon_{B}$
  measurements are effectively upper limits on $\epsilon_{B}$.

\subsection{Comparison Of Our Results On $\epsilon_{B}$ To Previous
  Studies}\label{KS_Test}
We performed a KS test between our optical $\epsilon_{B}$
measurements and the results from previous studies on $\epsilon_{B}$ 
(Figure \ref{literature_compilation_histograms}). For our $\epsilon_{B}$ results, we
used the optical $\epsilon_{B}$ measurements with $n = 1 \mbox{ cm}^{-3}$ and $p$ determined from
$\alpha_{O}$ (filled-in histogram in Bottom-Right panel of 
Figure \ref{optical_epsilon_B_measurement_histograms}; the
$\epsilon_{B}$ values are shown in Table \ref{optical_properties_table}).
The result of the KS test is that the null hypothesis is rejected. The P-value, which measures
the probability that the null hypothesis is still true, is $2.1 \times 10^{-9}$.
This result shows that the rejection of the null hypothesis is statistically significant.
It is not surprising that the null hypothesis was rejected. The distribution from the previous studies is very
inhomogeneous, with the values for $\epsilon_{B}$ being
drawn from many different studies with different methodologies. Also, comparing the
histogram in Figure \ref{literature_compilation_histograms}  to the filled-in
histogram in the Bottom-Right panel of Figure \ref{optical_epsilon_B_measurement_histograms}, 
a couple of significant differences can be seen. 
The range for the histogram of $\epsilon_{B}$ values found in the
literature is $ \sim 10^{-5}-10^{-1}$, whereas the
range for our $\epsilon_{B}$ results is
$\sim 10^{-8}-10^{-3}$. The mean and median values
for these two histograms are also significantly different. The mean (median)
value for the $\epsilon_{B}$ histogram from the literature, $6.3
\times 10^{-2}$ ($1.4 \times 10^{-2}$), is a factor $\sim 700$ ($\sim 600$)
times larger than the mean (median) $\epsilon_{B}$ value of the
histogram with our results, which is $9.5 \times 10^{-5}$ ($2.4 \times 10^{-5}$). 

One assumption that is commonly made in afterglow modeling studies is
equipartition between $\epsilon_{e}$ and $\epsilon_{B}$. As we
discussed in Section \ref{section_2}, the results for
$\epsilon_{e}$ from the literature and the results from recent
simulations of relativistic collisionless shocks support 
$\epsilon_{e} \sim 0.2$. From this result, many works assume 
$\epsilon_{B} \sim 10^{-2} - 10^{-1}$. However, there is no physical argument
to expect equipartition. Our distribution of $\epsilon_{B}$
upper limits and measurements, although wide, supports that there is
no equipartition between electron and magnetic energies because
none of the $\epsilon_{B}$ upper limits or measurements in our samples
has a value larger than 
$\epsilon_{B} \sim \mbox { few } \times 10^{-3}$. 
Another source of error that can lead to differences in
$\epsilon_{B}$ values is differences in the determination of the
spectral regime for the optical band. We took it to be between
$\nu_{i}$ and $\nu_{c}$, but it is also possible for the optical band
to be above $\nu_{c}$ at late times 
\citep[e.g.][]{panaitescu_and_kumar_2002,cenko_et_al_2010}. 
Another source for error is energy injection. We did
not consider energy injection as a source of error because only $3/35$
of the bursts in our optical sample show plateaus (and these plateaus
are short). The X-ray and optical light curves of many bursts show
plateaus and in these cases energy injection needs to be
considered. Also, errors in our determination of fluxes and times
from X-ray and optical light curves can also lead to small errors in $\epsilon_{B}$.
In summary, the main assumption we made when determining
$\epsilon_{B}$ is an efficiency of $ \sim 20\%$ in the conversion
of kinetic energy to gamma-ray energy, and we did not assume
equipartition between $\epsilon_{e}$ and $\epsilon_{B}$. Different
authors have made different assumptions that
can have a large effect on the results for $\epsilon_{B}$. 

Lastly, for a few bursts, we checked if our method of
  determining $\epsilon_{B}$ is consistent with the values determined
  for $\epsilon_{B}$ with other techniques. GRBs 980519 and 990123,
  discussed in the afterglow modelling study of
  \citealt{panaitescu_and_kumar_2002}, have optical light curves that
  decline as a power-law before the jet-break. The optical band for
  both of these bursts was determined to be in the spectral regime
  $\nu_{i} < \mbox{ 2 eV} < \nu_{c}$. Applying our technique to find
  a $\epsilon_{B}$ measurement for both of these bursts  and using the
 value of $n$ reported in \citealt{panaitescu_and_kumar_2002} for both
 of these bursts, we find that these bursts have $\epsilon_{B} \sim
 10^{-5}$, consistent with the results reported in
 \citealt{panaitescu_and_kumar_2002} for both of these bursts within a
 factor of a few. The small differences in $\epsilon_{B}$ values can
be accounted for by differences in the coefficients used for the
external-forward shock flux.

\section{GRBs In Common To Both Our X-ray And Optical Samples}\label{both_x-ray_and_optical}

\begin{figure}[t]
\centering
\includegraphics[scale=0.6]{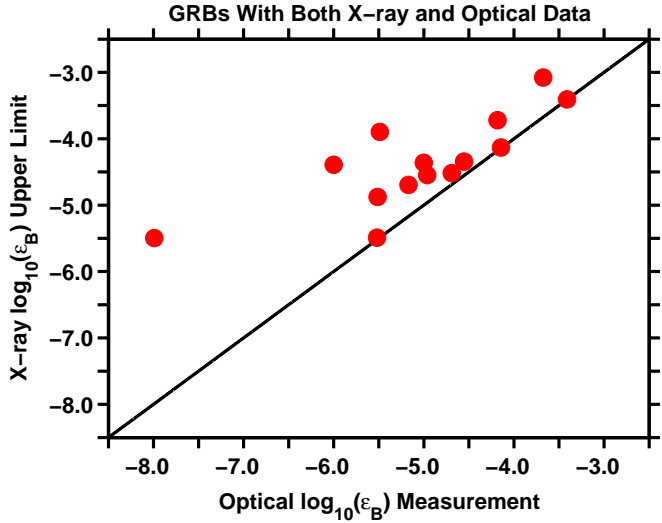}
\caption{Comparison of the $\epsilon_{B}$ upper limits
from X-ray data to the $\epsilon_{B}$ measurements from optical
data. The 14 dots correspond to the 14 GRBs that are both in our
X-ray and optical samples. The straight line indicates where the $\epsilon_{B}$
measurements are equal to the $\epsilon_{B}$ upper limits. \label{x_ray_and_optical_epsilon_B_comparison} }
\end{figure}

14 bursts we considered are both in our X-ray
and in our optical sample (GRB number is in bold in the first columns of 
Tables \ref{x-ray_properties_table} and \ref{optical_properties_table}). 
In this section, we verify for these bursts that the X-ray
$\epsilon_{B}$ upper limit is above the optical $\epsilon_{B}$ measurement. 
For the optical data, we will use the $\epsilon_{B}$ measurements with
$p$ determined from $\alpha_{O}$ and $n = 1 \mbox{ cm}^{-3}$
(filled-in histogram in the Bottom-Right panel of Figure \ref{optical_epsilon_B_measurement_histograms}). 
For the X-ray data, in this section, we will also consider $n =1 \mbox{ cm}^{-3}$
($s=0$) and we will use the value of $p$ determined from $\alpha_{O}$
\footnote{10 out of 14 of these bursts have optical data
    before 1000 seconds. For these 10 bursts, we can use the optical
    data to check if they satisfy the assumption we made in Section
    \ref{steep_decline_overview}, $t_{\mathrm{dec}} <
    t_{\mathrm{EoSD}}$. 9 of these 10 GRBs do satisfy this assumption;
    for the remaining GRB (080721), we are not able to check this assumption
    because the first optical observation (at 100 sec.) is after the
    end of the steep decline ($t_{\mathrm{EoSD}}=70$ sec).}. The 
comparison between the X-ray $\epsilon_{B}$ upper limits
and the optical $\epsilon_{B}$ measurements is shown in 
Figure \ref{x_ray_and_optical_epsilon_B_comparison}. This plot 
shows that all the X-ray $\epsilon_{B}$ upper limits 
are above the optical $\epsilon_{B}$ measurements. 

\section{$E$ vs. $\epsilon_{B}$ Correlation?}\label{bursts_with_density_and_E_vs_epsilon_B_correlation}

\begin{figure}[t]
\centering
\includegraphics[scale=0.6]{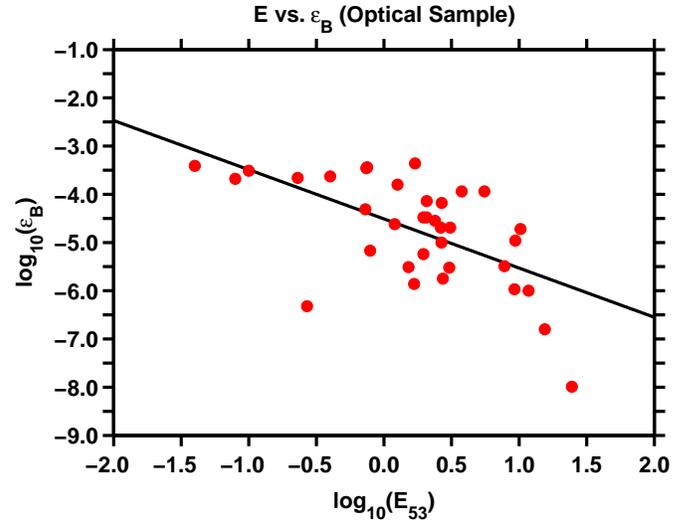}
\caption{We plot the values of $E$ and the
  measurements of $\epsilon_{B}$ to determine if they are correlated.
  The 35 points represent the GRBs in our optical sample and the straight line is the best fit
  line: $\mbox{log}_{10}(\epsilon_{B}) = -1.02 \mbox{log}_{10}(E_{53})
-4.51$, with the slope of the line being $-1.02 \pm 0.23$ and the
y-intercept of the line being $-4.51 \pm 0.16$. The correlation
coefficient of the fit is 0.62 and the P-value of the
correlation is $1.2 \times 10^{-4}$ (3.8$\sigma$ significance). The $\epsilon_{B}$ measurements 
are for $n = 1 \mbox{ cm}^{-3}$ and $p$ determined 
from $\alpha_{O}$ (shown in the filled-in histogram in the
Bottom-Right panel of
Figure \ref{optical_epsilon_B_measurement_histograms} and 
Table \ref{optical_properties_table}) and the values of $E$ were
determined by assuming an efficiency $\sim 20\%$ for all the GRBs in
our optical samples. \label{E_vs_epsilon_B_correlation} }
\end{figure}

We now use the $\epsilon_{B}$ measurements from our optical sample to determine if there is a correlation between 
$E$ and $\epsilon_{B}$. The 3 bursts studied in 
\citealt[][]{kumar_and_barniol_duran_2009,kumar_and_barniol_duran_2010}  have small
values of $\epsilon_{B}$, consistent with shock
compression of a seed magnetic field 
$B_{0} \sim \mbox{ few} \times 10 \mu \mbox{G}$.  
One property that distinguishes these 3 GRBs is that they were particularly
energetic, with 
$E_{\mathrm{iso}} ^{\gamma} \sim 10^{53} - 10^{54} \mbox{ ergs}$. 
Could the large energy intrinsic to these 3 bursts
explain why these 3 bursts have low $\epsilon_{B}$ values?
We investigate this possibility in Figure
\ref{E_vs_epsilon_B_correlation} by plotting the values of $E$ and
$\epsilon_{B}$ for all the bursts in our optical sample.

For the energy of each burst, we
assumed an efficiency of $\sim 20\%$ ($E = 5 E^{\gamma} _{\mathrm{iso}}$) 
in the conversion of the kinetic energy of the
jet to gamma-ray radiation; the observables involved in
calculating $E$ are the gamma-ray fluence and
$z$. For the $\epsilon_{B}$ measurements, we used the
values with $p$ determined from $\alpha_{O}$ and 
$n = 1 \mbox{ cm}^{-3}$. The observables involved in determining
$\epsilon_{B}$ are the observed specific optical flux and the time.  
In Figure \ref{E_vs_epsilon_B_correlation}, we also show the best fit
line\footnote{From Equation \ref{x-ray_constraint}, it may be expected that $E$ and
$\epsilon_{B}$ are correlated. This is not necessarily true because
each burst has a different value for the observed quantities $f_{\nu}$
and $t$. In addition, since $p$ was determined from $\alpha_{O}$, each
burst has a different $p$. We also checked if $E$ and $\epsilon_{B}$
were correlated when assuming a fixed $p=2.4$. From Equation
\ref{epsilon_B_afterglow_parameters}, we would expect the slope of the
line to be $-1.59$. The best fitting line for $p=2.4$ 
is $\mbox{log}_{10}(\epsilon_{B}) = -0.77 \mbox{log}_{10} (E_{53}) -
4.53$, with the slope of the line being $-0.77 \pm 0.24$ and the
y-intercept being $-4.53 \pm 0.16$. The slope of this line
is more than $3 \sigma$ away from $-1.59$, showing that there is no
expected correlation between $E$ and $\epsilon_{B}$ for our
methodology of determining $\epsilon_{B}$.}. This fit shows
that an increase in $E$ leads to a
decrease in $\epsilon_{B}$. The correlation coefficient of the
fit is $0.62$, indicating that there is a weak correlation between $E$
and $\epsilon_{B}$. A value of the correlation coefficient
close to 1 would indicate a strong correlation. The P-value of the
correlation is $1.2 \times 10^{-4}$ (3.8$\sigma$ significance), indicating that there is a
small probability that the correlation occurred by chance.

Although many points show large deviations from the best fit line, the scatter of the points
may be reduced or increased by the uncertainty in the afterglow
parameters. An error in the
efficiency would affect the values of $E$ and an error in
$\epsilon_{e}$, the efficiency, or $n$ would
affect the $\epsilon_{B}$ measurements (see Section \ref{X-ray_results}
for a discussion on how the $\epsilon_{B}$ measurements would be
affected when an error in a afterglow parameter is made).
It is possible that the uncertainty in 
$E$ and $\epsilon_{B}$ can reduce or increase  the scatter and make the correlation
between $E$ and $\epsilon_{B}$ stronger or weaker.


\section{Magnetic Field Amplification Factor For X-ray And Optical Results}\label{AF_X-ray}

In Sections \ref{section_3}-\ref{both_x-ray_and_optical}, we presented our results for the strength of
the magnetic field downstream of the shock front in terms of the afterglow parameter
$\epsilon_{B}$. If shock
compression was the only mechanism amplifying the ambient magnetic field 
(assuming a standard $B_{0} \sim \mbox{few } \mu \mbox{G}$ and a standard
$n = 1 \mbox{ cm}^{-3}$), then $\epsilon_{B} \sim 10^{-9}$ is expected.
Most of the bursts in our distributions of
$\epsilon_{B}$ upper limits and measurements have values
larger than $\epsilon_{B} \sim 10^{-9}$. These results
suggest that amplification of the magnetic field, in addition to shock compression, 
is needed to explain the afterglow
observations. In this section, we will present our results in terms of an amplification
factor, which quantifies the amplification that is needed, beyond
shock compression, to explain the observations. 

If shock compression were the
only mechanism amplifying the seed magnetic field $B_{0}$, then 
$B = 4 \Gamma B_{0}$. To quantify how much additional amplification of the
ambient magnetic field is needed, beyond shock compression, we
define the amplification factor, $AF$, as
\begin{equation}
AF \equiv \frac{B}{4 B_{0} \Gamma} .
\end{equation}
$AF$ is a constant that satisfies $AF \ge 1$ since $B \ge 4 B_{0}
\Gamma$. $AF = 1$ means than the observed $B$ is consistent with the
only amplification arising from seed magnetic field 
shock compression. The expression for $\epsilon_{B}$ is
$\epsilon_{B} = B^{2}/32 \pi m_{p} c^{2} n \Gamma^{2}$.
With the definition for $AF$, $\epsilon_{B}$ is
\begin{equation}
\epsilon_{B} = (AF)^{2} \times \frac{B_{0} ^{2}}{2 \pi
      n m_{p} c^{2}} . \label{epsilon_B_amp_factor}
\end{equation}
We note that $\epsilon_{B}$ is given by $(AF)^{2}$ times the
$\epsilon_{B}$ we would get if shock compression were the only
mechanism amplifying the magnetic field. 

We will now use Equation \ref{epsilon_B_amp_factor} and our previous
results for the X-ray $\epsilon_{B}$ upper limits to determine an upper limit on
$AF$.
In Section \ref{X-ray_results}, if we assumed a standard 
$n = 1 \mbox{ cm}^{-3}$, we were able to attain an upper limit on
$\epsilon_{B}$ for a constant density 
medium \footnote{We will only consider a constant density medium when
displaying the results for the amplification factor. We will show in 
Equation \ref{x_ray_amp_factor_constraint}
that $AF$ has a weak dependence on the density.}. We will refer to these 
$\epsilon_{B}$ upper limits as
$\overline{\epsilon_{B}} (p, n_{0}=1 )$. In the
notation $\overline{\epsilon_{B}} (p, n_{0}=1 )$, the bar over $\epsilon_{B}$
signifies that this is an upper limit on $\epsilon_{B}$, the
$p$ in the parenthesis shows that the $\epsilon_{B}$ upper limit depends on the value of $p$ we used,
and the $n_{0}=1$ shows that we assumed $n=1 \mbox{ cm}^{-3}$. With
this notation, we can keep the dependence of the $\epsilon_{B}$
upper limit on $n$ (see Equation \ref{x-ray_constraint}):
\begin{equation}
\epsilon_{B} < \frac{\overline{\epsilon_{B}} (p, n_{0}=1 )}{n_{0} ^{2/(p+1)}}. \label{previous_eps_B_upper_lim}
\end{equation}
Combining Equation \ref{epsilon_B_amp_factor}
and Equation \ref{previous_eps_B_upper_lim}, the expression for the upper
limit on $AF$ is
\begin{equation}
AF < \frac{1.0 \times 10^{4} [\overline{\epsilon_{B}} (p,
  n_{0}=1 )]^{1/2}}{B_{0, 10 \mu \mathrm{G}} n_{0} ^{(1-p)/(2p+2)}} \label{x_ray_amp_factor_constraint},
\end{equation}
where $B_{0, 10 \mu \mathrm{G}} \equiv B_{0} / 10 \mu \mbox{G} $.
One advantage to expressing the results of the magnetic field
downstream of the shock front in terms of $AF$ is
that $AF$ depends weakly on $n$. For a standard $p=2.4$, $AF \propto n^{0.21}$. On the other hand, there is a
strong dependence on $B_{0}$, $AF \propto B_{0} ^{-1}$.  

\subsection{Amplification Factor Upper Limit For Our X-ray Sample}\label{AF_upper_lim_results}

\begin{figure*}[t]
\begin{center}$
\begin{array}{c}
\begin{array}{cc}
\hspace{-10mm}
\includegraphics[scale=0.6]{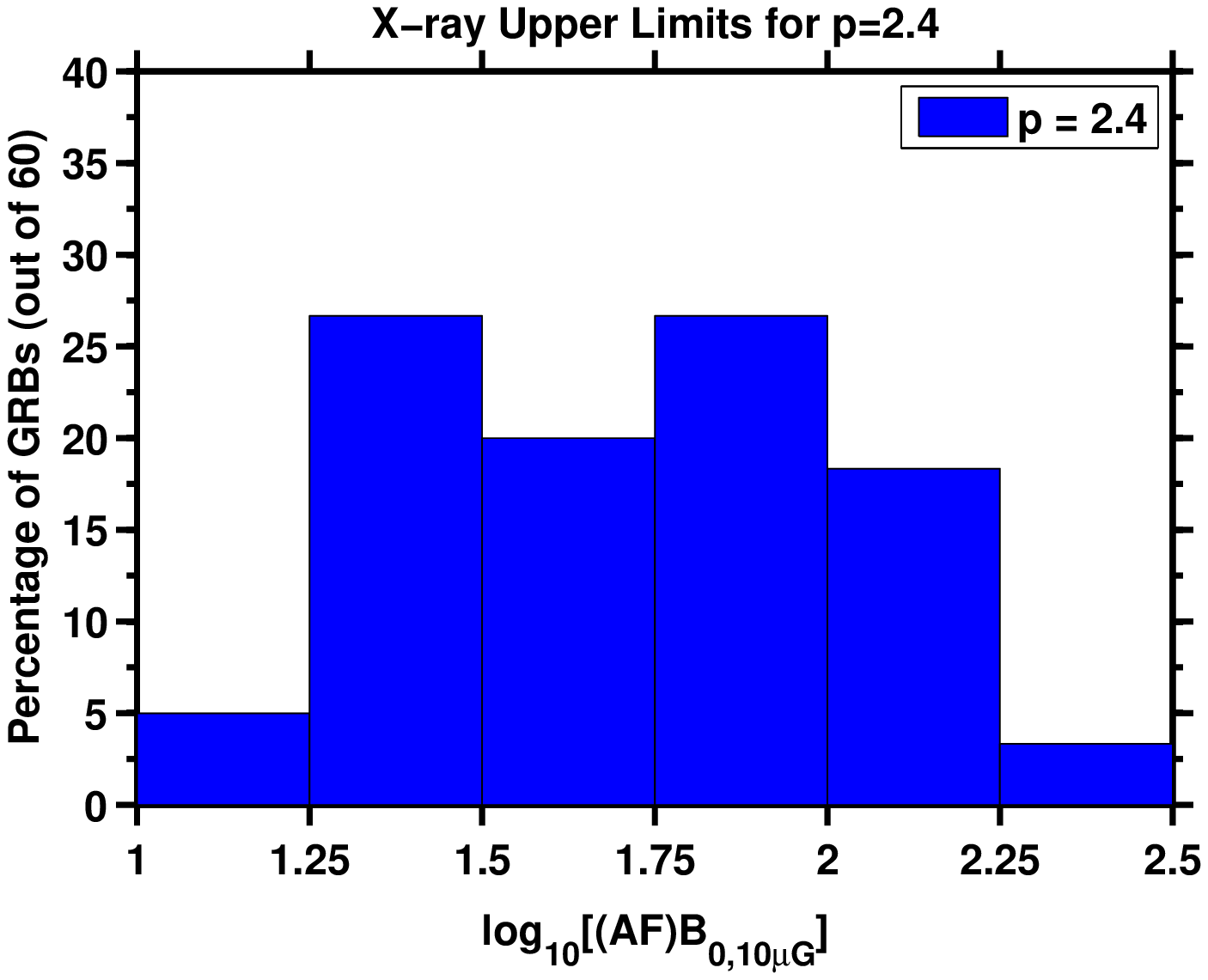} &
\hspace{0mm}
\includegraphics[scale=0.6]{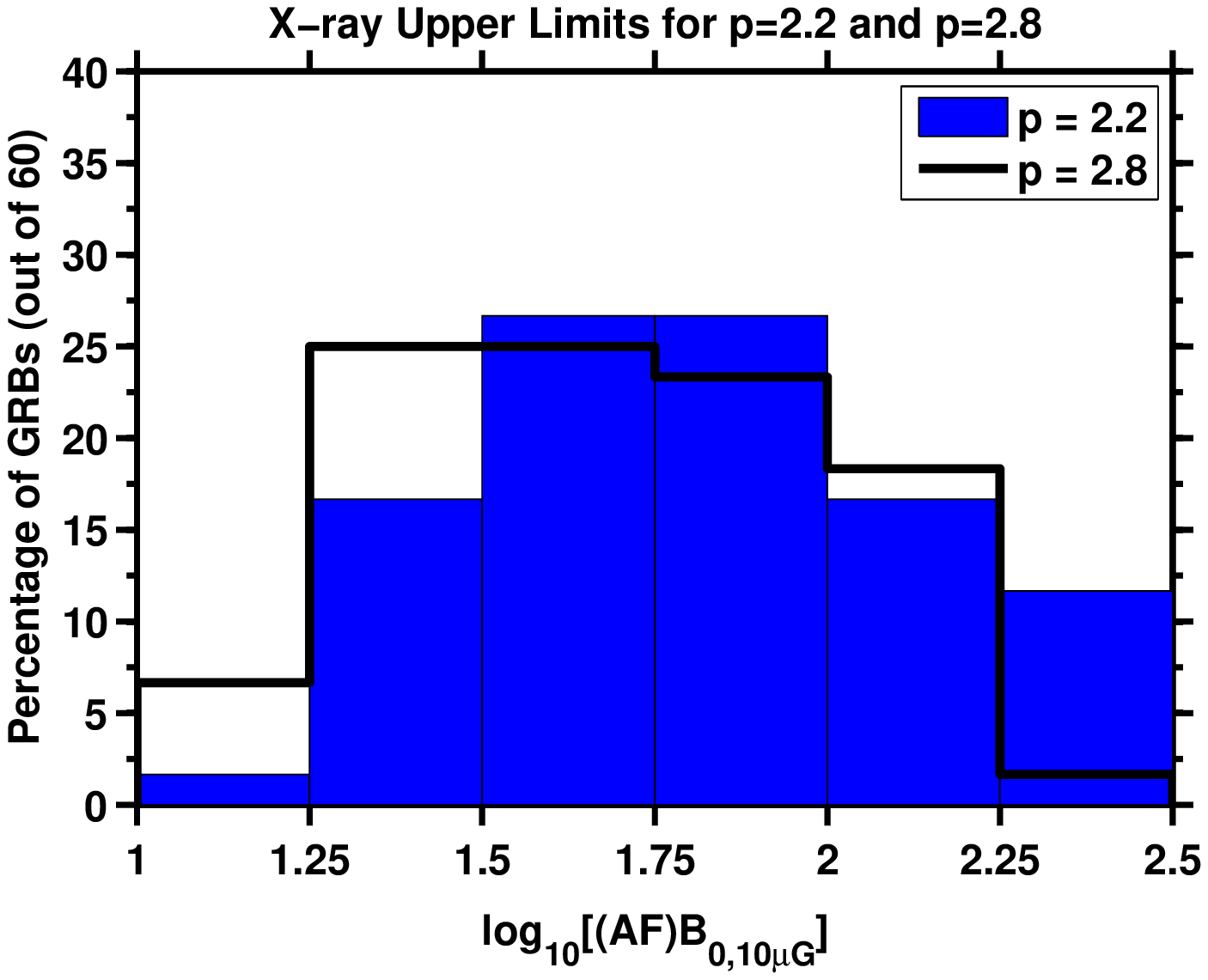}
\end{array}
\end{array}$
\end{center}
\caption{\textit{Left:} Upper limits on the quantity 
$(AF) B_{0, 10 \mu \mathrm{G}}$ for our X-ray sample assuming $p=2.4$. \textit{Right:} Upper
limits on the quantity $(AF) B_{0, 10 \mu \mathrm{G}}$ for our X-ray sample assuming $p=2.2$ and
$p=2.8$. A fixed $n=1 \mbox{ cm}^{-3}$ was assumed for all of the
histograms (the precise value of $n$ is unimportant since $AF$ has a
weak dependence on $n$, see Section \ref{AF_X-ray} and 
Equation \ref{x_ray_amp_factor_constraint}).  \label{x-ray_amplification_factor_upper_limits}}
\end{figure*}

We will now show the results for the $AF$ upper limits we obtained
from Equation \ref{x_ray_amp_factor_constraint}. Since the
amplification factor has a weak dependence on the density, we will
assume a standard $n_{0}=1$ when displaying the results for the $AF$
upper limits. When plotting the results for $AF$, we will keep the
dependence on $B_{0}$ and plot the quantity $(AF) B_{0, 10 \mu \mathrm{G}}$.
In the left and right panels of Figure \ref{x-ray_amplification_factor_upper_limits},
we show the upper limits on the quantity 
$(AF) B_{0, 10 \mu \mathrm{G}}$ for a fixed $p=2.4$ and $p=2.2$, 2.8, where the values of 
$\overline{\epsilon_{B}} (p, n_{0}=1 )$ used in
Equation \ref{x_ray_amp_factor_constraint} were shown in 
Top-Right, Top-Left, and Bottom panels of Figure
\ref{x-ray_epsilon_upper_limit_histograms}, respectively. For the
remainder of this section, we will assume $B_{0} = 10 \mu \mbox{G}$
when discussing the results for the $AF$ upper limits.

The mean and median values of the $AF$ upper limits are summarized
in Table \ref{AF_results_table}. The mean (median) $AF$ upper limits
range from $AF \sim 60-80$ ($AF \sim 40-60$). The $AF$
upper limit histograms show a wide distribution, with a range of 
$\sim 10$ to $\sim 300$. To determine if assuming a different value of $p$ has a
significant effect on the distribution of $AF$ upper limits, we
performed a KS test between the histograms in the right panel of
Figure \ref{x-ray_amplification_factor_upper_limits}. The KS
test confirmed the null hypothesis, leading us to conclude that the
$AF$ upper limit results are not sensitive
to the value of $p$ we assume.
\begin{table}[t]
\begin{small}
\begin{tabular}{| l | c c c c | }
\hline															
X-ray &  &  &  & \\
$(s=0)$	&	$p=2.2$	&	$p=2.4$	&	$p=2.8$	&  \\
\hline	
Mean	&	$ 84 $	&	$ 67 $	&	$ 62 $	&  \\
Median	&	$ 63 $	&	$ 51 $	&	$ 44 $	&  \\
\hline
\hline															
Opt. &  &  &  & \\
$(s=0)$	&	$p=2.2$	&	$p=2.4$	&	$p=2.8$	& $p$ from $\alpha_{\mathrm{O}}$ \\
\hline
Mean &	$ 130 $	&	$ 71 $	&	$ 36 $ & $ 71 $ \\
Median &	$ 100 $	&	$ 56 $	&	$ 23 $ & 	$ 48 $ \\
\hline
\end{tabular}
\end{small}
\caption{Mean and median $AF$
  values for the X-ray (upper limits on $AF$) and optical
  (measurements of $AF$) histograms shown in Figures \ref{x-ray_amplification_factor_upper_limits}
and \ref{optical_amplification_factor_upper_measurements}. All the labels are
the same as in Table \ref{epsilon_B_results_table}. A constant density
medium with $n = 1 \mbox{ cm}^{-3}$ (the amplification factor has a
weak dependence on the density, see Section \ref{AF_X-ray} 
and Equation \ref{x_ray_amp_factor_constraint}) and a
seed magnetic field $B_{0} = 10 \mu \mbox{G}$ were assumed
for all the bursts in our X-ray and optical samples. 
\label{AF_results_table}}
\end{table}

We now discuss how an error in each of the afterglow
parameters can affect our results for the $AF$ upper limits. For this
discussion, we will assume $p=2.4$. 
From Equation \ref{x_ray_amp_factor_constraint}, 
$\mbox{AF} \propto \sqrt{\overline{\epsilon_{B}} (p=2.4, n_{0}=1 )} B_{0} ^{-1}
n^{0.2}$ and from Equation \ref{epsilon_B_afterglow_parameters},
$\epsilon_{B} \propto \epsilon_{e} ^{-1.6} E^{-1.6} n^{-0.6}$. From these two expressions, 
$\mbox{AF} \propto \epsilon_{e} ^{-0.8} E^{-0.8} n^{0.2} B_{0} ^{-1}$. 
We note that compared to the $\epsilon_{B}$ upper
limit (Equation \ref{epsilon_B_afterglow_parameters}), the $AF$ upper
limit has a weaker dependence on $\epsilon_{e}$, $E$, and $n$. As we discussed at the end of Section
\ref{X-ray_results}, a likely error
in $\epsilon_{e}$ is a factor of $\sim 2$; this error in $\epsilon_{e}$
will translate into an error in the $AF$ upper limits by a factor of
only $\sim 2$. For the energy, we assumed an efficiency $\sim 20\%$ and
a likely error in the efficiency is a factor $\sim 2-3$ 
(see Section \ref{X-ray_results}); this error in the efficiency would lead to an error in the
$AF$ upper limits by only a factor $\sim 2$. One advantage to expressing 
the results of the magnetic field
downstream of the shock front in terms of $AF$
is that $AF$ has a very weak dependence on $n$. 
An error in $n$ by a factor $\sim 10^{3}$ (see Section \ref{X-ray_results}) from our assumed 
$n = 1 \mbox{ cm}^{-3}$ will only lead to an error in the $AF$ upper
limits by a factor $\sim 4$. The price to pay for a weak $n$ dependence is a linear dependence on
$B_{0}$, with $\mbox{AF} \propto B_{0} ^{-1}$. $B_{0}$ is an
uncertain parameter that likely varies from GRB environment to GRB
environment and it is the largest source of uncertainty for
$AF$. 

\subsection{Amplification Factor Measurement For Our Optical Sample}\label{AF_optical}

\begin{figure*}[t]
\begin{center}$
\begin{array}{c}
\begin{array}{cc}
\hspace{-10mm}
\includegraphics[scale=0.6]{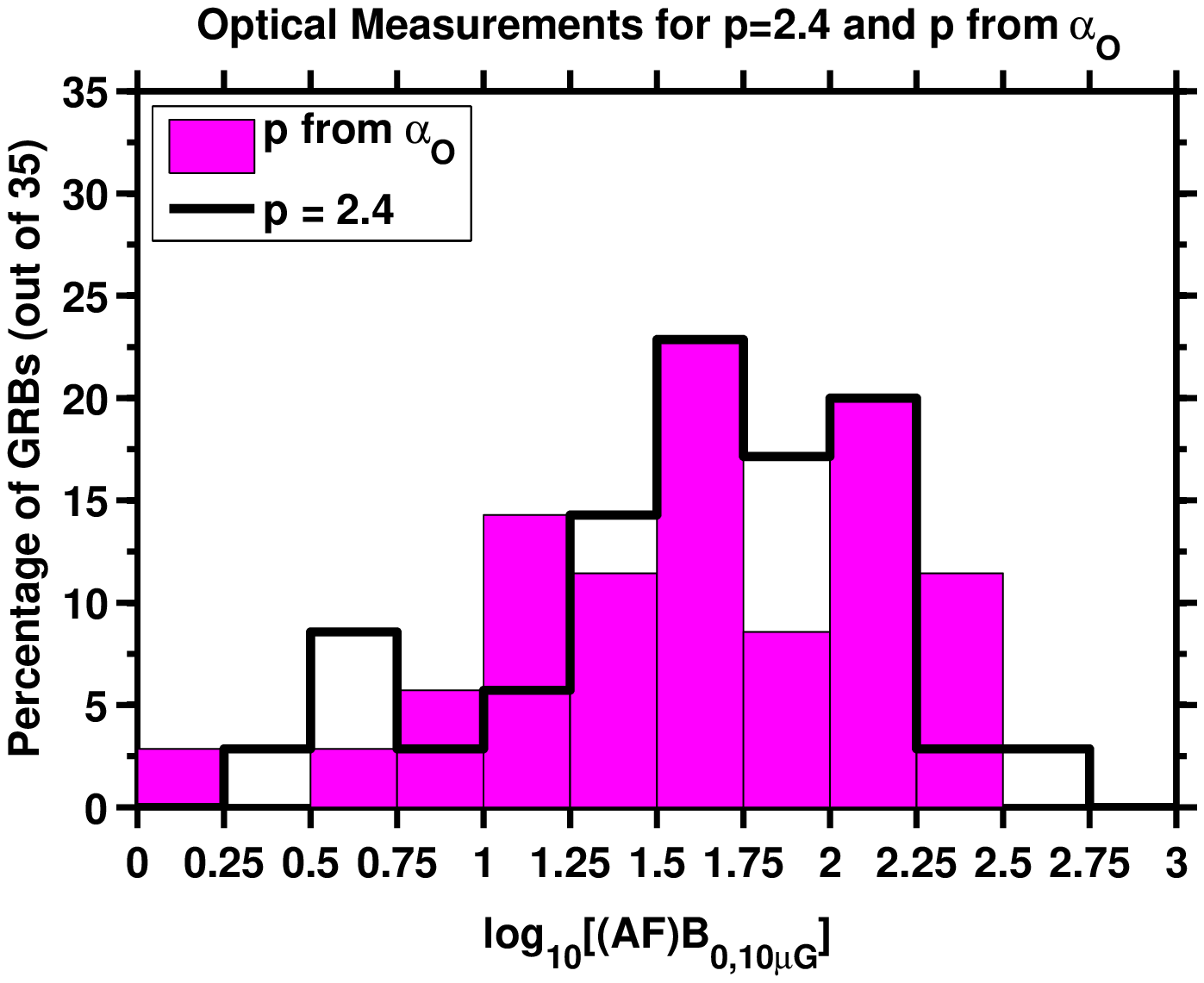} &
\hspace{0mm}
\includegraphics[scale=0.6]{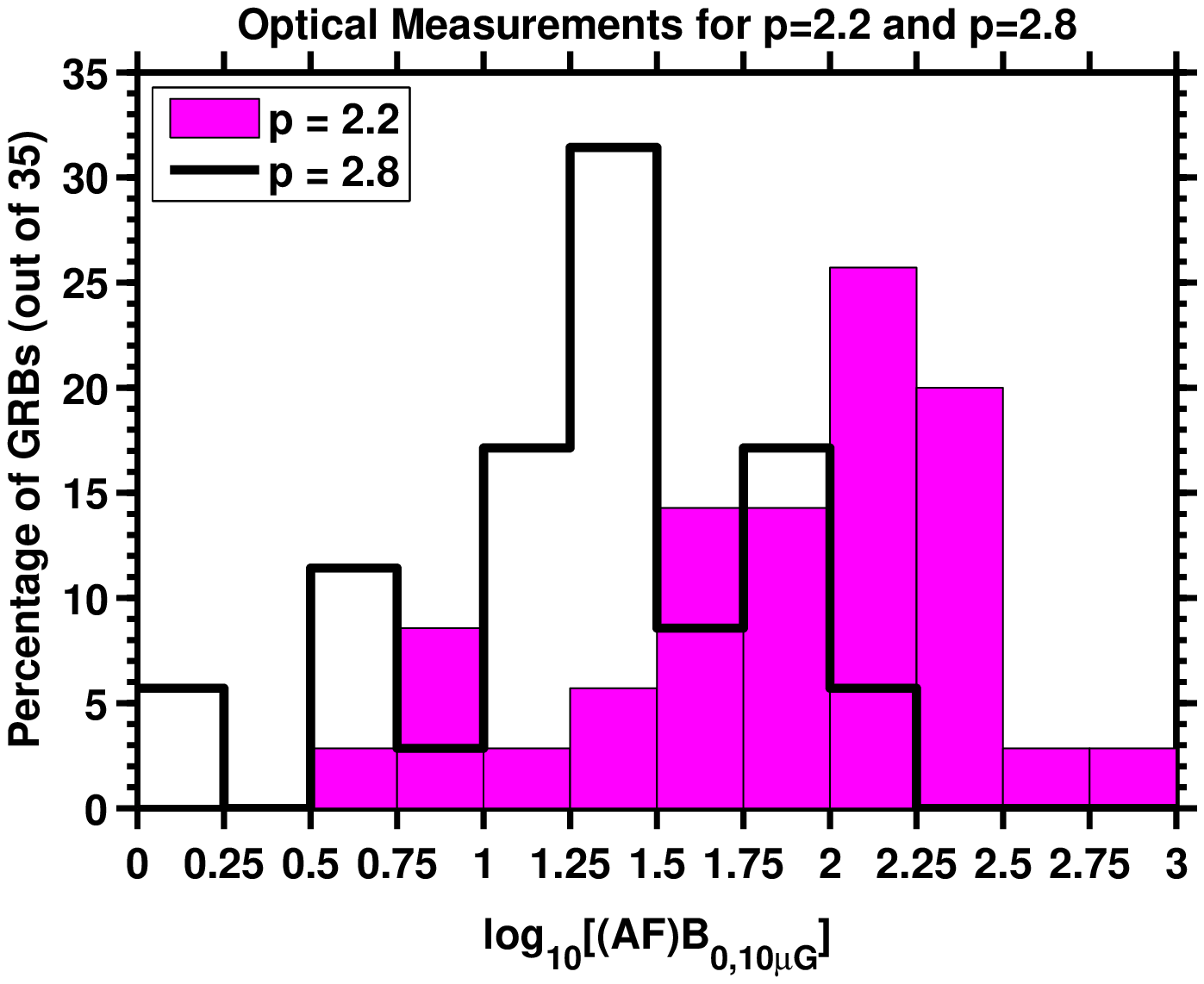}
\end{array}
\end{array}$
\end{center}
\caption{\textit{Left Panel:} The filled-in (un-filled) histogram
shows the measurements on the quantity $(AF) B_{0, 10 \mu \mathrm{G}}$ for $p$ calculated from $\alpha_{O}$
($p=2.4$). \textit{Right Panel:} The filled-in (un-filled) histogram
shows the measurements on the quantity $(AF) B_{0, 10 \mu \mathrm{G}}$
for $p=2.2$ ($p=2.8$). A fixed $n=1 \mbox{ cm}^{-3}$ was assumed for all of the
histograms (the precise value of $n$ is unimportant since $AF$ has a
weak dependence on $n$, see Section \ref{AF_X-ray} and 
Equation \ref{x_ray_amp_factor_constraint}). \label{optical_amplification_factor_upper_measurements}}
\end{figure*}

As we discussed in Section
\ref{optical_afterglow_parameter_assumption}, for our optical sample,
we found a measurement for $\epsilon_{B}$ instead of an upper
limit. This will allow us to determine a measurement for $AF$. To do
this, we will use Equation \ref{x_ray_amp_factor_constraint}, but in this case we have an
equality instead and we have $\epsilon_{B} (p, n_{0}=1 )$ instead of $\overline{\epsilon_{B}} (p,
  n_{0}=1 )$. The notation $\epsilon_{B} (p, n_{0}=1 )$ denotes the
$\epsilon_{B}$ measurements for our optical sample from
Section \ref{optical_epsilon_{B}_results} if we assume a standard 
$n =1 \mbox{ cm}^{-3}$. Also, as with the X-ray
sample, we only consider $s=0$ when calculating the
$AF$ measurements and assume a fixed $n = 1 \mbox{ cm}^{-3}$. 
In the left panel of Figure \ref{optical_amplification_factor_upper_measurements}, we show 
the results for the measurements on the quantity $(AF) B_{0, 10 \mu \mathrm{G}}$ for $p$ determined from
$\alpha_{O}$ and also assuming $p=2.4$. 
In the right panel of Figure \ref{optical_amplification_factor_upper_measurements}, we show the
measurements on the quantity $(AF) B_{0, 10 \mu \mathrm{G}}$ for $p=2.2$ and $p=2.8$. For the remainder of this
section, we will assume $B_{0} = 10 \mu \mbox{G}$ when discussing the
results for the $AF$ measurements. 

A summary of the mean and median $AF$ measurements for our optical sample is
shown in Table \ref{AF_results_table}. To determine if assuming a
standard $p=2.4$, as opposed to determining $p$ from $\alpha_{O}$ for
each burst, has a statistically significant effect on the distribution of $AF$
measurements, we performed a KS test between the two histograms in the
left panel of Figure \ref{optical_amplification_factor_upper_measurements}. The KS
test confirmed the null hypothesis. The mean (median) $AF$
measurements for the optical histograms range from 
$\sim 40$ to $\sim 130$ ($\sim 20$ to $\sim 100$). Compared to the
$AF$ upper limit histograms, the $AF$ measurement histograms show a
wider distribution, ranging from $AF \sim 1$ to $AF \sim
1000$. Also, since we used the same expression to determine the $AF$
upper limits and measurements (Equation
\ref{x_ray_amp_factor_constraint}), the discussion at the end of
Section \ref{AF_upper_lim_results} on how an error in one of the
afterglow parameters can affect the $AF$ upper limits also applies to
the $AF$ measurements\footnote{$\xi$ and $f$ also affect our $AF$ results. To account for
  $\xi$, since the $\epsilon_{B}$ upper limit$/$measurement is $\propto
\xi^{4(p-2)/(p+1)}$ and the $AF$ upper limit$/$measurement is 
$\propto (\overline{\epsilon_{B}} (p, n_{0}=1 ))^{1/2}$, 
$AF \propto \xi^{2(p-2)/(p+1)}$. To
account for $f$, since $AF \propto (\overline{\epsilon_{B}} (p, n_{0}=1
))^{1/2}$, $AF \propto f^{1/2}$. Thus, including $\xi$ and $f$ will
decrease the values of the $AF$ upper limits$/$measurements. 
Taking the lowest possible value for $f$, the $AF$ upper
limits$/$measurements can decrease by up to a factor $\sim 40$. 
This would make $\sim 50\%$($\sim 60\%$) of the bursts in our 
X-ray (optical) sample consistent with shock compression. In addition, 
as with the $\epsilon_{B}$ measurements, since $\xi$ and $f$ are less 
than unity, including these two parameters will mean that our $AF$
measurements will become upper limits on $AF$.}.

\section{Discussion And Conclusions}\label{conclusions_and_implications}

In this work, we presented a systematic  study on the magnetic field
downstream of the shock front for large samples of GRBs (60 in our
X-ray sample and 35 in our optical sample). We expressed the
strength of the downstream magnetic field in terms of both the
afterglow parameter $\epsilon_{B}$ and a amplification factor, denoted
by $AF$, which quantifies how much amplification, beyond shock
compression of the seed magnetic field, is needed to explain the
downstream magnetic field. This is the first time a large and
systematic study has been carried out to study $\epsilon_{B}$ and to
determine how much amplification of the seed magnetic field is
required by the observations. For our X-ray (optical) sample, we
determined an upper limit (measurement) for both $\epsilon_{B}$ and
$AF$. The upper limits on $\epsilon_{B}$ and $AF$ for our X-ray sample
were found from the constraint that the observed flux at the end of
the steep decline is greater than or equal to the external-forward
shock flux. This is a new method to constrain $\epsilon_{B}$
that relies on the steep decline emission, which has been observed by
\textit{Swift} for many GRBs. Our optical sample was restricted to
light curves that decline with $\alpha \sim 1$ from the early times
$\sim 10^{2} - 10^{3}$ sec, as expected for the external-forward shock
emission. We found the measurements for $\epsilon_{B}$ and $AF$ for
our optical sample from the condition that the observed flux is equal
to the external-forward shock flux. 

The condition used for our X-ray
(optical) sample was converted into an upper limit (measurement) on
the quantity $\epsilon_{B} n ^{2/(p+1)}$ for $s=0$ or $\epsilon_{B}
A_{*-1} ^{4/(p+1)}$ for $s=2$ by assuming a
$\sim 20\%$ efficiency in the conversion of kinetic energy to prompt
gamma-ray radiation. To find an upper limit (measurement) on
$\epsilon_{B}$ for our X-ray (optical) sample, we assumed a standard
$n = 1 \mbox{ cm}^{-3}$ ($A_{*} = 0.1$) for the density for
a constant density (wind) medium. A discussion on how the uncertainty
in the afterglow parameters affects our results for $\epsilon_{B}$ can
be found at the end of Section \ref{X-ray_results}. The largest source
of uncertainty for our results on $\epsilon_{B}$ is the density, since
the value of the density has been observed to vary over many orders 
of magnitude and its precise value is not known for each GRB 
(see Section \ref{X-ray_afterglow_parameter_assumptions}). For the
bursts that are both in our X-ray and optical samples, we also applied a
consistency check to make sure our results for $\epsilon_{B}$ are
correct (see Section \ref{both_x-ray_and_optical}).

From Table \ref{epsilon_B_results_table}, for a constant density (wind)
medium, most of the $\epsilon_{B}$ upper limit and measurement
histograms have a median value $\sim \mbox{ few} \times 10^{-5}$ 
($\sim \mbox{ few} \times 10^{-6}$). These results imply that half of
the bursts in both our X-ray and optical samples have a $\epsilon_{B}$
value $\sim \mbox{ few} \times 10^{-5}$ or lower. Assuming $n = 1 \mbox{ cm}^{-3}$ and 
$B_{0} \sim \mbox{few} \times \mu$G, shock compression is only able to produce
$\epsilon_{B} \sim 10^{-9}$. Although $\epsilon_{B} \sim 10^{-9}$ is 4
orders of magnitude lower than $\epsilon_{B} \sim 10^{-5}$,
$\epsilon_{B} \sim 10^{-5}$ is smaller by $\sim 2-4$ orders of
magnitude compared to the majority of previously reported
$\epsilon_{B}$ values (Figure \ref{literature_compilation_histograms}), which are 
$\epsilon_{B} \sim 10^{-3} - 10^{-1}$. Assuming $B_{0} \sim 10 \mu\mbox{G}$,
$\epsilon_{B} \sim \mbox{ few} \times 10^{-5}$ corresponds to 
$AF \sim 50$ (Equation \ref{x_ray_amp_factor_constraint}). Our result
of a median $\epsilon_{B} \sim \mbox{ few} \times 10^{-5}$ shows
that the majority of the bursts in our X-ray and optical samples only
require a weak amplification beyond shock compression, by a factor
$\sim 50$ or lower. 

The near equipartition $\epsilon_{B} \sim 0.01-0.1$ determined near the shock front
by theoretical studies and Particle-In-Cell (PIC) simulations
\citep{medvedev_and_loeb_1999,chang_et_al_2008,martins_et_al_2009,
keshnet_et_al_2009,lemoine_2013,sironi_et_al_2013} 
stands in contrast with our median results of $\epsilon_{B} \sim \mbox{ few} \times 10^{-5}$.
PIC simulations of relativistic collisionless shocks performed by
\citealt{chang_et_al_2008} and \citealt{keshnet_et_al_2009}
found that the magnetic field generated near the shock front decays with distance downstream of the shock front. 
\citealt{lemoine_2013} and \citealt{lemoine_et_al_2013} studied the effects that this
decaying magnetic field has on the shock
accelerated electrons radiating afterglow emission downstream of the shock front.
The main effect is that electrons with different Lorentz
factors cool in regions with different magnetic fields, with the higher (lower) energy
photons being emitted by electrons that are closer (further) from the
shock front \citep{lemoine_et_al_2013}. Considering the decay of the
downstream magnetic field,
\citet{lemoine_et_al_2013} modelled the afterglow data of 4 GRBs that have extended emission at energies 
$> 100 \mbox{ MeV}$ (detected by \textit{Fermi}-LAT) and also X-ray, optical, and radio data. Their 
afterglow modelling results for the X-ray, optical, and radio data
found $\epsilon_{B} \sim 10^{-6} - 10^{-4} $,
consistent with our results for the median $\epsilon_{B}$ upper limits
and measurements attained from X-ray and optical data.

Our next main result relates to the distribution of $\epsilon_{B}$
values. One property the $\epsilon_{B}$ values from the
literature shared with our optical $\epsilon_{B}$ measurements is that they both show
a wide distribution. The literature
compilation (Figure \ref{literature_compilation_histograms}) 
showed $\epsilon_{B} \sim 10^{-5} -10^{-1}$ and our optical
$\epsilon_{B}$ measurement histograms 
showed an even wider distribution, ranging from 
$\epsilon_{B} \sim 10^{-8} - 10^{-3}$
($\epsilon_{B} \sim 10^{-9} - 10^{-3}$) for a constant density (wind)
medium. One possibility we investigated to explain the wide
distribution of $\epsilon_{B}$ values
is whether bursts with smaller 
$\epsilon_{B}$ values are more energetic than
bursts with larger $\epsilon_{B}$ values 
(Figure \ref{E_vs_epsilon_B_correlation}). Although the bursts in 
our optical sample did show the trend that bursts with
larger $E$ have a smaller $\epsilon_{B}$, the correlation was
weak, with a correlation coefficient of $0.62$ and a P-value of
$1.2 \times 10^{-4}$ (3.8$\sigma$ significance).

Another possibility to explain the wide distribution of
$\epsilon_{B}$ relates to the uncertainty in the environmental
parameters ($B_{0}$ and $n$) in the medium surrounding GRBs. 
One possibility to explain bursts with values of
$\epsilon_{B} \sim 10^{-5} - 10^{-3}$, under the interpretation of weak
amplification beyond shock compression, is 
that they occurred in environments with particularly high seed magnetic
fields. Since $\epsilon_{B} \propto B_{0} ^{2}$, an increase in $B_{0}$ by an order of
magnitude will lead to an increase in $\epsilon_{B}$ by two orders of
magnitude. A weak amplification beyond shock
compression of $AF \sim 50$ (as inferred for the bursts with
$\epsilon_{B} \sim \mbox{ few} \times 10^{-5}$) and a strong
$B_{0} \sim 10 \mu \mbox{G} - 100 \mu \mbox{G}$ yields
$\epsilon_{B} \sim 10^{-5} - 10^{-3}$ 
(see Equation \ref{x_ray_amp_factor_constraint}). We note that for this estimate
we assumed $n = 1 \mbox{ cm}^{-3}$, but as we discussed in 
Section \ref{AF_X-ray}, the amplification factor has a weak dependence
on the density. This possibility, that the larger values of $\epsilon_{B}$ may be
explained by bursts going off in environments with particularly strong
$B_{0}$, will be discussed further in Barniol Duran 2013 (in preparation).

In addition to many bursts in our optical sample having
particularly large $\epsilon_{B} \sim 10^{-5}- 10^{-3}$, 
there were also some bursts in our optical sample that have particularly 
low $\epsilon_{B}$ values. For $p$
determined from $\alpha_{O}$ and $n = 1 \mbox{ cm}^{-3}$ 
(Bottom-right panel of Figure \ref{optical_epsilon_B_measurement_histograms}), 
GRBs 071025 and 071112C have 
$\epsilon_{B} \sim 10^{-7}$ and GRB 080607 has 
$\epsilon_{B} \sim 10^{-8}$. Assuming $n = 1 \mbox{ cm}^{-3}$ and 
$B_{0} = 10 \mu \mbox{G}$, these bursts with 
$\epsilon_{B} \sim 10^{-8} - 10^{-7}$ are consistent with shock 
compression of a seed magnetic field of
a few 10 $\mu \mbox{G}$ being the only amplification necessary to explain the
observations.  

Lastly, we mention that a similar
conclusion for radio observations of supernova remnants (SNRs) was
reached in \citet{thompson_et_al_2009}. Like GRB afterglow emission, SNR
emission arises from a blastwave interacting with the surrounding medium, but 
at non-relativistic speeds. \citet{thompson_et_al_2009} found that starburst galaxies
have strong ambient magnetic fields $\sim 1$ mG and they concluded that shock
compression of this strong seed magnetic field is enough to explain the radio emission from SNRs. On the
other hand, for normal spiral galaxies with ambient magnetic fields
$\sim 5-10 \mu$G, they concluded that additional amplification beyond
shock compression by a factor $\sim 3-7$ was necessary to explain the radio emission from
SNRs.


\acknowledgments
This work made use of data supplied by the UK Swift Science
Data Centre at the University of Leicester. 
RS dedicates this work to his friend, Jossue Colato, thanks Patrick
Crumley and Roberto Hern\'andez for helpful discussions, and Kevin
Gately for his help with obtaining the values of $\epsilon_{e}$ and
$\epsilon_{B}$ from the literature. This work has been funded in part
by NSF grant ast-0909110. RBD was supported by an ERC advanced grant
(GRB) and by the I-CORE Program of the PBC and the ISF (grant 1829$/$12).
%



\appendix

\section{$\epsilon_{e}$ And $\epsilon_{B}$ Values From The Literature}\label{literature_values_appendix}

In this Appendix, we show a table of the $\epsilon_{e}$ and $\epsilon_{B}$ values we used to make the
histograms in Figure \ref{literature_compilation_histograms}. {For the
first GRBs with high quality afterglow data, 
different works 
(e.g. \citealt{wijers_and_galama_1999,chevalier_and_li_1999,chevalier_and_li_2000,li_and_chevalier_et_al_2001,harrison_et_al_2001,yost_et_al_2003})
have presented afterglow modeling studies on the same GRBs. 
In Table \ref{epsilon_e_and_epsilon_B_table}, for these early afterglow
studies, we show the results from 
\citealt{panaitescu_and_kumar_2001a,panaitescu_and_kumar_2001b,panaitescu_and_kumar_2002} since they have the largest
afterglow modeling compilations. It is important to note that
in some cases, different groups determined significantly different values for the afterglow
parameters 
(e.g. GRB 970508
\citealt{wijers_and_galama_1999,chevalier_and_li_2000,frail_et_al_2000,panaitescu_and_kumar_2002}, 
GRB 000418 \citealt{berger_et_al_2001,panaitescu_and_kumar_2002}).

\begin{center}
\begin{longtable}{|c|c|c|c|}
\caption{$\epsilon_{e}$ and $\epsilon_{B}$ Values From The Literature \label{epsilon_e_and_epsilon_B_table}}\\
\hline
GRB & 
$\epsilon_{e}$  &
$\epsilon_{B}$  &
Ref. \\
\hline
\endfirsthead
\caption{$\epsilon_{e}$ and $\epsilon_{B}$ Values From The Lit. (Continued)}\\
\hline
GRB & 
$\epsilon_{e}$  &
$\epsilon_{B}$  &
Ref. \\
\hline
\endhead
\hline 
\endfoot
970508	&	$ 0.62 $	&	$ 0.10 $	&	[1]	\\
980329	&	$ 0.12 $	&	$ 0.17 $	&	[2]	\\
980519	&	$ 0.25 $	&	$ (3.5^{+32} _{-2.3}) \times 10^{-5} $	&	[1]	\\
980703	&	$ 0.14 $	&	$ 4.6 \times 10^{-4} $	&	[3]	\\
990123	&	$ 0.59 $	&	$ (7.4^{+23} _{-5.9}) \times 10^{-4} $	&	 [1]	\\
990510	&	$ > 0.3 $	&	$ 6 \times 10^{-3} $	&	[4]	\\
991208	&	$ 0.32 $	&	$ 2.1 \times 10^{-2} $	&	[1]	\\
991216	&	$ 0.4 $	&	$ 2 \times 10^{-2} $	&	[4]	\\
000301C	&	$ 0.4 $	&	$ 7 \times 10^{-2} $	&	[4]	\\
000926	&	$ 0.35 $	&	$ (6.5^{+1.5} _{-1.1}) \times 10^{-2} $	&	 [1]	\\
010222	&	$ > 0.3 $	&	$ 2 \times 10^{-4} $	&	[4]	\\
011211	&	$ 0.22 $	&	$ 5.0 \times 10^{-4} $	&	[5]	\\
020405	&	$ 0.1$	&	$ 0.3 $	&	[6]	\\
020813	&		&	$ 4.0 \times 10^{-4} $	&	[5]	\\
021004	&	$ 0.21 $	&	$ 2 \times 10^{-4} $	&	[7]	\\
030226	&	$ 0.11 $	&	$ 2.5 \times 10^{-4} $	&	[5]	\\
030329	&	$ 0.16 $	&	$ 0.10 $	&	[8]	\\
050416A	&	$ 0.2 - 0.333 $	&	$ 0.072 - 0.333  $	&	[9]	\\
050820A	&	$ 0.14^{+0.02} _{-0.01} $	&	$ 0.013^{+0.003} _{-0.001} $	&	[10]	\\
050904	&	$ 0.02 $	&	$ 0.015 $	&	[11]	\\
051022	&	$ 0.0681^{+0.3951} _{-0.0348} $	&	$ (8.02^{+28.18} _{-7.17}) \times 10^{-3} $	&	[12]	\\
051221A	&	$ 0.24 - 0.333 $	&	$ 0.12 - 0.333 $	&	[13]	\\
060418	&	$ 0.06^{+0.01} _{- 0.02} $	&	$ 0.15^{+0.14} _{-0.01} $	&	[10]	\\
070125	&	$ 0.27^{+0.03} _{-0.01} $	&	$ 0.0277^{+0.0044} _{-0.0075} $	&	[14]	\\
080129	&	$ 0.4 $	&	$ 5 \times 10^{-2} $	&	[15]	\\
080319B	&	$ 0.11 \pm 0.01 $	&	$ 0.33 $	&	[10]	\\
080928	&	$ 0.165 $	&	$ (2.5^{+16} _{-2.4}) \times 10^{-4} $	&	[16]	\\
090323	&	$ 0.070^{+0.005} _{-0.005} $	&	$ 0.0089^{+0.0007} _{-0.0018} $	&	[17]	\\
090328	&	$ 0.11^{+0.06} _{-0.01} $	&	$ 0.0019^{+0.0004} _{-0.0008} $	&	[17]	\\
090423	&	$ 0.28 $ 	&	$ 1.6 \times 10^{-4} $	&	[18]	\\
\hline
\caption{\small{ In this table, we show all of the
$\epsilon_{e}$ and $\epsilon_{B}$ values, determined in previous
afterglow modelling studies, that we
were able to find in the literature. These
values are plotted in the histograms in Figure \ref{literature_compilation_histograms}. In the
column labeled Ref. we
give the reference where we found each value of $\epsilon_{e}$ and $\epsilon_{B}$.
Reference Legend: $[1]=$ \citet{panaitescu_and_kumar_2002},
$[2]=$ \citet{yost_et_al_2002},
$[3]=$ \citet{panaitescu_and_kumar_2001b},
$[4]=$ \citet{panaitescu_and_kumar_2001a},
$[5]=$ \citet{panaitescu_2005},
$[6]=$ \citet{berger_et_al_2003a},
$[7]=$ \citet{bjornsson_et_al_2004},
$[8]=$ \citet{berger_et_al_2003b},
$[9]=$ \citet{soderberg_et_al_2007},
$[10]=$ \citet{cenko_et_al_2010},
$[11]=$ \citet{frail_et_al_2006},
$[12]=$ \citet{rol_et_al_2007},
$[13]=$ \citet{soderberg_et_al_2006}
$[14]=$ \citet{chandra_et_al_2008},
$[15]=$ \citet{gao_2009},
$[16]=$ \citet{rossi_et_al_2011},
$[17]=$ \citet{cenko_et_al_2011},
$[18]=$ \citet{chandra_et_al_2010} }}
\end{longtable}
\end{center}

\end{document}